\documentclass[%
reprint,
superscriptaddress,
amsmath,amssymb,
aps,
prx,
]{revtex4-2}

\usepackage{graphicx}
\usepackage{braket}
\usepackage{booktabs}
\usepackage{amsthm}
\usepackage[hidelinks]{hyperref}

\newtheorem{definition}{Definition}[]
\newtheorem{theorem}{Theorem}[]
\newtheorem{lemma}[theorem]{Lemma}
\newtheorem{corollary}{Corollary}[theorem]

\begin{document}

\title{Fourier extensions for matrix-function block encodings with error-independent subnormalization bounds}

\author{Peter Brearley}
\email{peter.brearley@manchester.ac.uk}
\affiliation{Centre for Quantum Science and Engineering, University of Manchester, UK}
\affiliation{Department of Mechanical and Aerospace Engineering, University of Manchester, UK}
\author{Thomas L.~Howarth}
\affiliation{Department of Aeronautical and Automotive Engineering, Loughborough University, UK}
\author{Ben Adcock}
\affiliation{Department of Mathematics, Simon Fraser University, Canada}

\begin{abstract}
     Block encodings of non-unitary matrix functions are central to quantum numerical linear algebra. Hamiltonian simulation is a natural input model for Hermitian matrices, but accurate block encodings often incur large subnormalization. We decouple the accuracy from the subnormalization by formulating the matrix-function block encoding as a Fourier-extension approximation problem, yielding a linear combination of unitaries for Hermitian matrix inputs. Fourier extensions approximate non-periodic functions by a Fourier series on a larger periodic domain, creating redundant coefficients that can be optimized for their absolute sum, and hence subnormalization. The coefficients may be chosen for optimal subnormalization with algebraic convergence, by tuning the subnormalization bound for increasing rates of exponential convergence, or by Sobolev-regularized fitting to accommodate more general spectral sets. Fourier-extension block encodings apply to eigenvalue transforms of Hermitian matrices, or to odd singular-value transforms of general matrices, including as a quantum linear systems algorithm.
\end{abstract}

\maketitle

\section{Introduction}
\label{sec:introduction}

Problems across scientific computing can often be formulated in terms of matrix functions $f(\mathcal H)$ acting on vectors $\boldsymbol\psi$. Classical algorithms are tasked with computing this efficiently at a large scale on parallel computing infrastructure. On a quantum computer, an equivalent task is to efficiently prepare a quantum state proportional to $f(\mathcal H)\ket{\psi}$, given an amplitude encoding $\ket{\psi}$ of $\boldsymbol\psi$. Quantum computers are well-known to offer a potentially exponential algorithmic advantage for functions of the form $f(\mathcal H) = e^{i\mathcal H}$ when $\mathcal H = \mathcal H^\dagger$, as this represents the dynamics of a closed quantum system~\cite{Lloyd1996}. Exponential speedups are also attainable for other non-unitary $f(\mathcal H)$ under certain conditions, particularly when $f$ is smooth~\cite{Harrow2009, Subramanian2019}. Since the pioneering work of \citeauthor{Harrow2009}~\cite{Harrow2009} in demonstrating a quantum linear systems algorithm (QLSA) for the case $f(\mathcal H) = \mathcal H^{-1}$, known as the HHL algorithm, quantum algorithms for non-unitary matrix functions have been the subject of much attention~\cite{Childs2017, Subramanian2019, Takahira2022, An2023, Motlagh2024, Low2025, Wang2026, An2026laplace, Morales2025}, which we now discuss. 

Matrix functions $f(\mathcal H)$ transform the eigenvalues $\lambda_j$ of $\mathcal H$ in its spectral decomposition. A Hermitian matrix $\mathcal H \in \mathbb C^{N\times N}$ has the spectral decomposition $\mathcal H = \sum_{j=1}^N \lambda_j\ket{u_j}\!\bra{u_j}$, where $\ket{u_j}$ are the corresponding eigenvectors. The transformed matrix is
\begin{equation}
    f(\mathcal H) = \sum_{j=1}^N f(\lambda_j) \ket{u_j}\!\bra{u_j},
    \label{eq:spectral_basis}
\end{equation}
for any $f:\mathbb R\to \mathbb C$. Thus, even for Hermitian $\mathcal H$, the matrix function $f(\mathcal H)$ may be non-Hermitian when $f$ is complex-valued.

To solve the linear systems problem by preparing $\mathcal H^{-1}\ket{\psi}$, the HHL algorithm~\cite{Harrow2009} uses a decomposition of $\ket{\psi}$ in the spectral basis of $\mathcal H$ in Eq.~\eqref{eq:spectral_basis}, then applies quantum phase estimation to a Hamiltonian simulation of $\mathcal H$ to encode approximations of $\lambda_j$. Matrix functions can then be applied by probabilistic non-unitary transformations of the eigenvalue register. HHL approximates the inverse function by controlled rotations, but other functions can be applied if they can be efficiently approximated by a quantum circuit. After reversing quantum phase estimation, the quantum state is approximately proportional to $f(\mathcal H)\ket{\psi}$. HHL is limited by the first-order algebraic convergence of quantum phase estimation, and by the ability to represent $f$ using controlled rotation gates. It generalizes to odd singular-value transforms of non-Hermitian $A = U\Sigma V^\dagger$ by embedding $A$ into a Hermitian matrix $\mathcal H(A)$ with eigenvalues $\pm \Sigma$, known as a Hermitian dilation. Matrix inversion is the singular-value transform $A^{-1}=V\Sigma^{-1}U^\dagger$.

\citeauthor{Childs2017}~\cite{Childs2017} proposed a QLSA that avoids quantum phase estimation by applying $\mathcal H^{-1}$ directly as a linear combination of Hamiltonian-simulation unitaries $e^{-i\mathcal H t}$. This retains the Hamiltonian-simulation input model of HHL, but allows for improved precision dependence in approximating $f(\mathcal H)$ through Fourier or Chebyshev approximations~\cite{Childs2017}. Linear combination of unitaries (LCU) is a general strategy for constructing block encodings, which represent non-unitary matrices as blocks of unitary operators~\cite{Childs2012}. The concept behind block encoding has long been studied~\cite{Gingrich2004, Terashima2005}, though it has only been formalized in more recent works~\cite{Low2019, Gilyen2019}. LCU recasts the block-encoding problem into one of finding a decomposition
\begin{equation}
    f(\mathcal H) \approx \sum_j \beta_j U_j, \qquad \alpha\equiv\sum_j |\beta_j|,
    \label{eq:alpha_definition}
\end{equation}
where $\alpha \ge \|f(\mathcal H)\|-\epsilon$ is the subnormalization and $\beta_j\in \mathbb{C}$ are the LCU coefficients. High subnormalization increases the postselection or amplitude-amplification cost, and excessively high subnormalization may render the computation indistinguishable from noise~\cite{Cai2023}. Finding accurate LCU decompositions with small coefficient $\ell^1$-norms is therefore essential for producing high-quality block encodings in the LCU framework.

For condition number $\kappa$, the scalar approximation problem of \citeauthor{Childs2017}~\cite{Childs2017} is to express $x^{-1}$ on the gapped spectral domain $[-1,-\kappa^{-1}]\cup[\kappa^{-1},1]$ as an LCU while keeping the coefficient $\ell^1$-norm small. They use the integral representation
\begin{equation}
    x^{-1} = \frac{i}{\sqrt{2\pi}} \int_0^\infty \! {\rm d}y \int_{-\infty}^{\infty} \! {\rm d}z\,z e^{-z^2/2} e^{-ixyz},
    \label{eq:childs_inverse_lcu}
\end{equation}
which is obtained by writing $x^{-1}=\int_0^\infty g(xy)\,{\rm d}y$ with the odd function $g(y)=y\exp(-y^2/2)$, which satisfies $\int_0^\infty g(y)\,{\rm d}y=1$, and then substituting the Fourier transform of $g$. This Gaussian choice ensures rapid decay in both the original and Fourier variables, so that truncating and discretizing the integral yields a finite LCU with polylogarithmic precision dependence. The resulting subnormalization scales as $\alpha = O(\kappa\sqrt{\log[\kappa/\epsilon]})$, which has a factor of $\sqrt{\log[\kappa/\epsilon]}$ above the optimal $O(\kappa)$. \citeauthor{Childs2017}~\cite{Childs2017} also developed a Chebyshev-based QLSA, replacing $x^{-1}$ by the polynomial $(1-[1-x^2]^b)/x$, which converges exponentially to $x^{-1}$ on the gapped spectral domain and admits an exact finite Chebyshev expansion. This improves the polylogarithmic precision dependence, but relies on a quantum-walk implementation of Chebyshev polynomials~\cite{Szegedy2004, Childs2010} rather than Hamiltonian simulations. The Chebyshev-based framework was later studied in the context of other smooth $f$~\cite{Subramanian2019}.

Both the HHL algorithm~\cite{Harrow2009} and the Fourier-LCU approach~\cite{Childs2017} transform the eigenvalues of a Hermitian matrix using Hamiltonian simulation as the input model. A possible means of extending this to eigenvalue transforms of general $A$ is through contour integrals~\cite{Takahira2022}, which express matrix functions as a linear combination of matrix inverses, although with subnormalization governed by potentially large resolvent norms. Laplace-transform representations reduce suitable eigenvalue transforms to linear combinations of non-unitary matrix exponentials~\cite{An2026laplace}, which may then be individually expressed as linear combinations of Hamiltonian simulations~\cite{An2023, An2026quantum}. This is restricted to stable half-plane spectra and to functions admitting suitable Laplace-transform representations~\cite{An2023, An2026quantum, An2026laplace}. Beyond the Hamiltonian-simulation input model, quantum singular value transformation (QSVT), and more recent extensions to eigenvalue transforms~\cite{Low2026eigenvalue}, have emerged as leading frameworks for quantum numerical linear algebra~\cite{Gilyen2019}, unifying several algorithms including QLSAs~\cite{Martyn2021}. QSVT assumes access to a block encoding of the input matrix, with the required number of accesses scaling linearly with the polynomial degree of the target approximation. This is a powerful approach to matrix functions when efficient block encodings are available, although the construction of such block encodings can be difficult and often dominates the practical cost~\cite{Nibbi2024, Lapworth2025}. Efficient block-encoding circuits are known for many structured operators~\cite{Wan2021, Sunderhauf2024}, but current general-purpose constructions typically suffer from large subnormalization~\cite{Schlimgen2021, Suri2023, Li2024, Bharadwaj2025}. QSVT faces operational challenges such as restrictions on the achievable polynomials and the often-intensive classical preprocessing stage of computing the phase angles that depend nontrivially on the target polynomial~\cite{Dong2021, Motlagh2024}.

In this work, we formulate block encodings of Hermitian matrix functions as a Fourier-extension approximation problem, improving the subnormalization properties compared to existing approaches~\cite{Harrow2009, Childs2017, Subramanian2019} and easing generalization to analytic $f$ on the spectral interval. The main idea is to approximate the function $f$ on the spectral interval using a Fourier series $f_m(x) = \sum_{k=-m}^mc_ke^{ikx}$ on a larger periodic domain, known as a Fourier extension~\cite{Huybrechs2010}. The redundancy of the Fourier basis on the restricted interval allows its coefficients to be chosen in a number of different ways depending on the objective. We give three coefficient-selection strategies that each bound the subnormalization independently of the error. Section~\ref{sec:optimal_subnormalization_on_extended_domains} gives LCU decompositions with optimal subnormalization, achieved by prescribing a reflected extension of $f$ outside of the fitted interval. Since the reflections are not smooth, the endpoint convergence is only algebraic, scaling as $O(m^{-1})$. Section~\ref{sec:exponential_convergence_with_bounded_subnormalization} then explicitly trades the subnormalization bound beyond the optimal value for increasing rates of exponential convergence.

\begin{theorem}
\label{thm:optimal_lcu_subnormalization}
    Let $\mathcal H = \mathcal H^\dagger\in\mathbb C^{N\times N}$, and let $f:\mathbb R\to \mathbb C$ satisfy the endpoint-reflection hypotheses of Lemma~\ref{lem:convex_endpoint_reflection_saturation}. Reflected-extension Fourier LCU prepares an $(\alpha, n_{\rm a}, \epsilon)$-block encoding of $f(\mathcal H)$ with optimal subnormalization
    \begin{equation}
        \alpha = \|f(\mathcal H)\|-\epsilon.
        \label{eq:optimal_subnormalization_f}
    \end{equation}
\end{theorem}

Theorem~\ref{thm:optimal_lcu_subnormalization} uses the definition of a block encoding given in Definition~\ref{def:block_encoding} of the following section. 

Our final strategy in Section~\ref{sec:sobolev_regularised_extensions} finds coefficients as the solution of a regularized approximation problem. We control the uniform $L^\infty$ approximation error (and hence the matrix spectral error) and coefficient $\ell^1$-norm using Sobolev norms, which are smooth while retaining the respective guarantees on the objectives. This allows for near-exponential convergence down to a target tolerance $\epsilon_{\rm t}$, with subnormalization $\alpha_m^{\rm reg}\leq C_f\|f(\mathcal H)\|$. The constant $C_f$ is independent of the target tolerance and is $O(1)$ for many common functions, including identity, exponential, and inverse functions, as shown in Section~\ref{sec:application_to_various_functions}.

\begin{theorem}
    \label{thm:sobolev_complexity}
    Let $\mathcal H=\mathcal H^\dagger\in\mathbb C^{N\times N}$, let $f:\mathbb R\to\mathbb C$ admit a holomorphic extension to a complex neighborhood of the fitted spectral set, and fix any $\sigma>1$. Sobolev-regularized Fourier LCU gives an $\epsilon$-accurate block encoding of $f(\mathcal H)$ with subnormalization $\alpha\leq C_f\|f(\mathcal H)\|$, where $C_f\geq1$ is defined in Eq.~\eqref{eq:Cf_extension_constant} and is independent of $m$ and $\epsilon$. Preparing a state proportional to $f_m(\mathcal H)\ket{\psi}$ requires
    \begin{equation}
        O\!\left(\frac{C_f\|f(\mathcal H)\|}{\|f_m(\mathcal H)\ket{\psi}\|_2}\log\!\log[\epsilon^{-1}]\right)
        \label{eq:regularized_hs_calls}
    \end{equation}
    uses of the Hamiltonian simulation algorithm, and if $\mathcal H$ is $s$-sparse, requires
    \begin{equation}
        O\!\left(\frac{C_f\|f(\mathcal H)\|}{\|f_m(\mathcal H)\ket{\psi}\|_2}s\log^\sigma[\epsilon^{-1}]\log N\right)
        \label{eq:regularized_complexity}
    \end{equation}
    two-qubit gates.
\end{theorem}

Since $C_f = O(1)$ for inversion, the subnormalization scales optimally as $O(\|A^{-1}\|)=O(\kappa)$. General non-Hermitian $A$ may be inverted via a Hermitian dilation, as used in the HHL algorithm~\cite{Harrow2009}. Formulated as a QLSA, we require $O(\kappa[\log\kappa+\log\!\log(\kappa\epsilon^{-1})])$ uses of the Hamiltonian simulation algorithm and $O(s\kappa^2\log^\sigma[\kappa\epsilon^{-1}]\log N)$ two-qubit gates. Compared to the Fourier-LCU QLSA of \citeauthor{Childs2017}~\cite{Childs2017}, this removes the $\sqrt{\log(\kappa/\epsilon)}$ factor from the subnormalization scaling. Both approaches are $\kappa^2$ in the gate complexity, although the precision dependence is improved from containing the factor $\log^5(\kappa/\epsilon)$ to $\log^\sigma(\kappa/\epsilon)$ for any $\sigma>1$, which may be arbitrarily close to $1$. QLSAs using other input models have achieved linear dependence on the condition number~\cite{Costa2022, Dalzell2024}. The QLSA constructions in Theorems~\ref{thm:optimal_lcu_subnormalization} and \ref{thm:sobolev_complexity} have an optimal and optimal-scaling subnormalization, respectively, so have optimal state-preparation dependence~\cite{Low2026linear}.

In Section~\ref{sec:lcu_decompositions_from_Fourier_series}, we introduce how Fourier series of $f$ yield LCU decompositions of $f(\mathcal H)$ for Hermitian matrix inputs. We also use this section to define much of the notation and conventions used throughout this work. In the context of Fourier extensions, the three proposed coefficient-selection strategies are given in Sections~\ref{sec:optimal_subnormalization_on_extended_domains}, \ref{sec:exponential_convergence_with_bounded_subnormalization}, and~\ref{sec:sobolev_regularised_extensions}. Section~\ref{sec:optimal_subnormalization_on_extended_domains} contains the proof of Theorem~\ref{thm:optimal_lcu_subnormalization}. In Section~\ref{sec:application_to_various_functions}, we compare the strategies numerically on several common functions. Section~\ref{sec:quantum_circuit} shows the standard LCU circuit construction, and how the required number of Hamiltonian-simulation calls can be reduced exponentially by exploiting that our unitaries are the same Hamiltonian evolution over different times. Section~\ref{sec:resource_requirements} discusses the resource requirements for each approach, including the proof of Theorem~\ref{thm:sobolev_complexity}. Section~\ref{sec:discussion} concludes the work.

\section{LCU decompositions from Fourier series}
\label{sec:lcu_decompositions_from_Fourier_series}

When implemented as a quantum circuit, the LCU decomposition of a non-unitary matrix function in Eq.~\eqref{eq:alpha_definition} prepares a block encoding, which we define as follows. 

\begin{definition}[Block encoding]
    An $(n + n_{\rm a})$-qubit unitary $U$ is an $(\alpha,n_{\rm a},\epsilon)$-block encoding of an $n$-qubit operator $A$ if $\| A - \alpha(\bra{0}^{\otimes n_{\rm a}} \otimes I)U(\ket{0}^{\otimes n_{\rm a}} \otimes I) \| \le \epsilon$.
    \label{def:block_encoding}
\end{definition}

In Definition~\ref{def:block_encoding} and throughout this work, matrix norms are spectral norms denoting the largest singular value. Applying $U$ to a quantum state with $n_{\rm a} = 1$ ancilla qubit approximates
\begin{equation}
    U(\ket{0}\otimes \ket{\psi}) = \ket{0}\otimes \frac{A\ket{\psi}}{\alpha} + \ket{1}\otimes \ket{{\rm junk}}
    \label{eq:block_encoding_transformation}
\end{equation}
up to error $\epsilon/\alpha$. This applies the desired non-unitary transformation $A$ to the $n$-qubit register $\ket{\psi}$ when the ancilla qubit is $\ket{0}$, and a `junk' transformation when the ancilla qubit is $\ket{1}$. The subnormalization $\alpha$ is a central property of block encodings as it determines the relative amplitudes of $A\ket{\psi}$ and $\ket{{\rm junk}}$ within the quantum state. The subnormalization must be at least the spectral norm $\alpha \ge \|A\|-\epsilon$ since the singular values of $U$ cannot exceed~1.

LCU has been applied in solvers for linear systems~\cite{Childs2017}, differential equations~\cite{Berry2017}, open quantum systems~\cite{Schlimgen2021}, imaginary time evolution~\cite{Chowdhury2017}, and fluid dynamics~\cite{Sanavio2024}. Many useful LCU block encodings exploit structure in the input matrix~\cite{Wan2021, Sunderhauf2024}. For example, it is straightforward to decompose $s$-sparse circulant matrices into a linear combination of $s$ unitaries~\cite{Wan2021}, which can be extended to various boundary conditions with additional pairs of unitaries that correct the boundary values~\cite{Over2025}. The requirement for highly structured matrices limits the applicability of LCU, since operators in the most demanding applications may not have such regular structures~\cite{Asanovic2006}. Without any structural assumptions about $A$, we rely on general LCU decompositions. A frequent approach approximates $A$ by a linear combination of four unitaries~\cite{Wu1994}, which has been applied to various applications~\cite{Schlimgen2021, Li2024, Bharadwaj2025}. Although the decomposition is $O(1)$ in the required number of unitaries, the subnormalization scales algebraically in the target error as $\alpha = O(\|A\|^{3/2}\epsilon^{-1/2})$. This necessitates high subnormalization to bring the block encoding to within a reasonable tolerance. Furthermore, it does not readily extend to encoding matrix functions $f(\mathcal H)$. 

A non-periodic function $f(x)$ can be approximated with a Fourier series $f_m(x)$ on a full period $x\in\mathbb T$ by a periodic continuation, where $\mathbb T = [-\pi,\pi)$. The Fourier series is defined by
\begin{equation}
    f_m(x) = \!\!\sum_{k=-m}^m c_k e^{ikx} = a_0 + \!\sum_{k=1}^m \left[a_k \sin(kx) + b_k \cos(kx)\right],
    \label{eq:fourier_series}
\end{equation}
where $c_k\in\mathbb{C}$ are the Fourier coefficients,  $c_0 = a_0$, and $c_{\pm k} = (b_k \mp ia_k)/2$ for $k=1,\dots,m$. In general, $a_0, a_k, b_k\in\mathbb{C}$, although when $f$ is real, $a_0,a_k, b_k\in\mathbb{R}$. By the functional calculus $x=\mathcal H$, Eq.~\eqref{eq:fourier_series} is an LCU decomposition when $\mathcal H=\mathcal H^\dagger$. When $f$ is periodic and analytic, the Fourier-LCU decomposition $f_m(\mathcal H)$ converges exponentially. If $f$ is a trigonometric polynomial, it is represented exactly by a finite Fourier series, which applies to smooth periodic filters. However, most matrix functions of interest are non-periodic, including the most basic identity setting of $f(\mathcal H) = \mathcal H$.

For $f_m(\mathcal H)$ to yield an LCU decomposition of the input matrix $\mathcal H$, we require a Fourier series of the identity function $f(x)=x$. Its periodic (sawtooth) continuation is a sine series expressed in terms of complex exponentials
\begin{equation}
    x = i\sum_{k=1}^m \frac{(-1)^{k}}{k}(e^{ikx} - e^{-ikx})+O(m^{-1}),
    \label{eq:sawtooth_continuation}
\end{equation}
with Fourier coefficients decaying as $a_k = O(k^{-1})$. This gives a diverging subnormalization $\alpha_m = \sum_{k=1}^m |a_k| = \Theta(\log m)$. Applying Eq.~\eqref{eq:sawtooth_continuation} to $x=\tau \mathcal{H}$ yields the LCU decomposition
\begin{equation}
    \mathcal{H} = \frac{i}{\tau}\sum_{k=1}^m \frac{(-1)^{k}}{k}(e^{ik\tau\mathcal{H}} - e^{-ik\tau\mathcal{H}}) + O (\|\mathcal H\|m^{-1}),
    \label{eq:sawtooth_lcu}
\end{equation}
where $\tau < \pi/\|\mathcal H\|$ may be chosen to scale the spectrum $\operatorname{spec}(\mathcal H) \subseteq [\lambda_{\rm min}, \lambda_{\rm max}]$ to within the fitted interval $\operatorname{spec}(\tau\mathcal H) \subset \mathbb T$, where $\lambda_{\rm min}$ and $\lambda_{\rm max}$ are the spectral endpoints. Since the Fourier series at the domain endpoints evaluates to $f_m(\pm\pi) = 0$, taking $\tau < \pi/\|\mathcal H\|$ avoids these points and the nearby Gibbs oscillations. The error of the block encoding in Definition~\ref{def:block_encoding} depends on the $L^\infty$-norm of the Fourier extension, which is $\epsilon_m = O(m^{-1})$ away from the endpoints, so the matrix spectral error is $\epsilon=O(\|\mathcal H\|m^{-1})$.

Assuming that $\tau$ is close to $\pi/\|\mathcal H\|$, the subnormalization is 
\begin{equation}
    \alpha_m^{\rm saw}\! = \frac{2}{\pi} \|\mathcal H\|\! \sum_{k=1}^m \frac{1}{k} = O(\|\mathcal H\|\log m) = O\!\left(\!\|\mathcal H\|\log \frac{\|\mathcal H\|}{\epsilon}\!\right)\!.
\end{equation}
Despite the sawtooth continuation converging only as $\epsilon_m = O(m^{-1})$, the subnormalization grows logarithmically in the target precision. This serves as the base case from which we improve the convergence to exponential and bound the subnormalization independently of the error using Fourier extensions.

The LCU decomposition generalizes to non-Hermitian matrices $A$ via the Hermitian dilation,
\begin{equation}
    \mathcal{H}(A) =
    \begin{pmatrix}
        0 & A^\dagger \\
        A & 0
    \end{pmatrix},
    \label{eq:hermitian_dilation}
\end{equation}
which requires supplementing the quantum state by an additional qubit such that
\begin{equation}
    \mathcal H(A)(\ket{0}\otimes \ket{\psi}) = \ket{1}\otimes A\ket{\psi}.
\end{equation}
This provides a route to block encoding singular-value transforms of general matrices. If $A = U\Sigma V^\dagger$ is a singular value decomposition, then the eigenvalues of $\mathcal H(A)$ are $\pm \Sigma$, so $f$ acts on the singular values of $A$. Decomposing $f = f_{\rm e} + f_{\rm o}$ into its even and odd parts, where $2f_{\mathrm e}(x) = f(x)+f(-x)$ and $2f_{\mathrm o}(x) = f(x)-f(-x)$, gives
\begin{equation}
    f(\mathcal H(A))=
    \begin{pmatrix}
        V f_{\mathrm e}(\Sigma)V^\dagger &  V f_{\mathrm o}(\Sigma) U^\dagger\\
        U f_{\mathrm o}(\Sigma)V^\dagger & U f_{\mathrm e}(\Sigma) U^\dagger
    \end{pmatrix}.
    \label{eq:general_svt_block_structure}
\end{equation}
The off-diagonal blocks encode the odd part of the transform, while the diagonal blocks encode the even part. This is useful since many matrix functions of practical interest are odd. If $f$ is odd, then $f_{\mathrm e}=0$ so Eq.~\eqref{eq:general_svt_block_structure} prepares a block encoding of the odd singular-value transforms $U f(\Sigma)V^\dagger$ and $V f(\Sigma)U^\dagger$. This applies to the identity and inversion functions in particular, since $U\Sigma V^\dagger=A$ and $V\Sigma^{-1}U^\dagger=A^{-1}$ for nonsingular $A$~\cite{Harrow2009}.

Fourier extensions approximate the target function on a subinterval $\Omega_\eta$ of the full period $\mathbb{T}$, where $\eta>1$ is the domain extension factor. When $f$ is defined throughout $\mathbb{T}$, we define the subinterval as
\begin{equation}
    \Omega_\eta = [-\pi/\eta,\pi/\eta], \qquad \eta>1.
    \label{eq:omega_eta}
\end{equation}
For a Hermitian matrix $\mathcal H$ with spectrum $\operatorname{spec}(\mathcal H)\subseteq[\lambda_{\rm min},\lambda_{\rm max}]$, rescaling $\mathcal H$ so that its spectral interval is mapped exactly onto the fitted interval reduces the unused fitted domain and improves the error and coefficient-prefactor behavior. Let
\begin{equation}
    \mu = \frac{\lambda_{\rm max} + \lambda_{\rm min}}{2}, \quad
    \delta = \|\mathcal H-\mu I\| = \frac{\lambda_{\rm max} - \lambda_{\rm min}}{2}
    \label{eq:mu_delta_definition}
\end{equation}
be the spectral midpoint and half-width, respectively. For $\delta>0$, we define the scaled matrix
\begin{equation}
    \mathcal G = \tau(\mathcal H-\mu I),
    \qquad
    \tau = \frac{\pi}{\eta\delta},
    \label{eq:G_tau_definition}
\end{equation}
such that $\operatorname{spec}(\mathcal G)\subseteq\Omega_\eta$. The case of $\delta=0$ is trivial, since then $\mathcal{H}=\mu I$ and $f(\mathcal H)=f(\mu)I$. The fitted function is
\begin{equation}
    g(x)=f\!\left(\mu+\frac{x}{\tau}\right) = f\!\left(\mu+\frac{\eta\delta}{\pi}x\right),
    \quad x\in\Omega_\eta,
    \label{eq:g_definition}
\end{equation}
such that $f(\mathcal H)=g(\mathcal G)$. Thus, if $g_m$ is a Fourier approximation to a periodic extension of $g$ as in Eq.~\eqref{eq:fourier_series}, then the corresponding matrix approximation is $f_m(\mathcal H) = g_m(\mathcal G) = \sum_{k=-m}^m c_k e^{ik\mathcal G}$. Since $\mathcal G$ is Hermitian when $\mathcal H$ is Hermitian, each $e^{ik\mathcal G}$ is unitary, and may be expressed in terms of Hamiltonian simulations of the input matrix as $e^{ik\mathcal G}=e^{-ik\tau\mu}e^{ik\tau\mathcal{H}}$. The rescaling therefore only contributes a known phase to the Hamiltonian-simulation unitaries. The constant part of the LCU decomposition $c_0I$ is now represented exactly, with the non-constant part being governed by the spectral half-width $\delta$ rather than by the full scale $\|\mathcal H\|$.

Finally, we distinguish the various error definitions used throughout this work. In Definition~\ref{def:block_encoding}, $\epsilon$ is the error of the block encoding, satisfying
\begin{equation}
    \epsilon = \|f(\mathcal H)-f_m(\mathcal H)\| \leq \|g-g_m\|_{L^\infty(\Omega_\eta)} = \epsilon_m.
    \label{eq:spectral_error}
\end{equation}
It equals $\epsilon_m$ when $\mathcal H$ has an eigenvalue at the location of the maximum error. We use $\epsilon_{\psi}$ to refer to the error in the final state, given by
\begin{equation}
    \epsilon_\psi = \left\|\frac{f(\mathcal H)\ket{\psi}}{\|f(\mathcal H)\ket{\psi}\|_2} - \frac{f_m(\mathcal H)\ket{\psi}}{\|f_m(\mathcal H)\ket{\psi}\|_2}\right\|_2 \leq \frac{2\epsilon}{\|f(\mathcal H)\ket{\psi}\|_2},
    \label{eq:epsilon_psi}
\end{equation}
and exclude the trivial case of $\|f(\mathcal H)\ket{\psi}\|_2 = 0$ throughout. We use $\epsilon_{\rm HS} \leq \epsilon_\psi/N_{\rm HS}$ to refer to the precision required of an individual Hamiltonian simulation, where $N_{\rm HS}$ is the required number of Hamiltonian simulations. The Sobolev-regularized algorithm of Theorem~\ref{thm:sobolev_complexity} uses a target error $\epsilon_{\rm t}$, which it converges towards and maintains with increasing $m$. Regarding subnormalization, $\alpha_m = \sum_{k=-m}^m |c_k|$ refers to the subnormalization evaluated from the Fourier-coefficient $\ell^1$-norm with truncation order $m$, and $\alpha$ refers to the subnormalization of block encodings more generally.

\section{Optimal subnormalization with algebraic convergence}
\label{sec:optimal_subnormalization_on_extended_domains}

The process of constructing LCU decompositions of matrix functions from Fourier series has been introduced, so we now focus on optimizing their performance using extended domains. This section prioritizes subnormalization, while the following section prioritizes exponential convergence.

\subsection{Triangular extension of the identity function}
\label{sec:triangular_extension_of_the_identity_function}

We first consider the identity function $f(x)=x$ for block encoding general $A$, before generalizing the norm-saturation mechanism to other $f$. With the notation of Eqs.~\eqref{eq:G_tau_definition} and \eqref{eq:g_definition}, taking $\eta=2$ gives $\tau=\pi/(2\delta)$ and $\operatorname{spec}(\mathcal G)\subseteq[-\pi/2,\pi/2]$. The fitted scalar function is $g(x)=\mu+x/\tau$.

Closing the non-constant part outside the interval by a straight line gives the triangular wave, which has sine-series coefficients $a_k = (4/[\pi k^2])\sin(k\pi/2)$. The Fourier series now approximates a continuous function, with the discontinuity passed to its first derivative at $x=\pm \pi/2$. The rate of coefficient decay therefore improves by one order to $|a_k|=O(k^{-2})$, which naturally bounds the coefficient $\ell^1$-norm. The LCU decomposition is
\begin{equation}
    \mathcal{H} = \mu I + \frac{\delta}{i\pi} \sum_{k=1}^m a_k\!\left(e^{ik\mathcal G}-e^{-ik\mathcal G}\right) + O(\delta m^{-1}),
    \label{eq:regular_triangular_lcu}
\end{equation}
with subnormalization
\begin{equation}
    \alpha_m^{\rm refl} = |\mu|+\frac{8\delta}{\pi^2}\sum_{k=1}^m\frac{|\sin(k\pi/2)|}{k^2}.
\end{equation}
Only the odd modes contribute, so
\begin{equation}
    \alpha_\infty^{\rm refl} = |\mu|+\frac{8\delta}{\pi^2}\sum_{\substack{k\geq1\\ k\ \mathrm{odd}}}\frac{1}{k^2} = |\mu|+\delta = \|\mathcal H\|,
    \label{eq:regular_triangular_optimal_subnormalization}
\end{equation}
since the sum over odd positive integers is $\pi^2/8$. The regular triangular extension achieves the optimal subnormalization. If we were to take $\tau<\pi/(2\delta)$, the spectrum of $\mathcal G$ lies strictly inside the fitted interval. This evaluates the triangular Fourier series away from the derivative discontinuities, which improves the error convergence by one order to $O(\delta m^{-2})$, but the subnormalization bound then exceeds $\|\mathcal H\|$.

The decomposition extends to general matrices $A$ using the Hermitian dilation in Eq.~\eqref{eq:hermitian_dilation}. Since the spectrum of $\mathcal H(A)$ is already centered, $\mu=0$ and $\delta=\|A\|$, so the scaled formulation offers no benefit. For general non-Hermitian matrix inputs, the triangular extension gives $\alpha_m^{\rm refl}= \|A\|-\epsilon$ with largest spectral error $\epsilon=O(\|A\|m^{-1})$.

\subsection{Application to other functions}

The triangular extension is an example of a more general norm-saturation mechanism. Any Fourier extension $g_m$ has coefficient $\ell^1$-norm, and therefore asymptotic LCU subnormalization $\alpha_\infty$, at least as large as the maximum magnitude of $g$ on the fitted interval. The lower bound is saturated when the Fourier modes are phase-aligned at a maximizer of $|g|$. We show that convex endpoint reflections provide a useful sufficient condition for this phase alignment, allowing functions such as $f(\mathcal H)=\mathcal H$, $\mathcal H^{-1}$, and $e^{\mathcal H}$ to be block encoded with optimal subnormalization. In this section, we consider the approximation interval $\Omega_\eta$ defined in Eq.~\eqref{eq:omega_eta}.

\begin{lemma}[Norm saturation criterion]
\label{lem:norm_saturation_criterion}
    Let $g:\Omega_\eta\to\mathbb C$, and let $\sum_{k \in \mathbb{Z}}c_{k}e^{ikx}$ be any $2\pi$-periodic extension of $g$ such that $\{c_k\}_{k\in\mathbb Z}\in\ell^1$. Then
    \begin{equation}
        \sum_{k\in\mathbb{Z}}|c_{k}| \geq \|g\|_{\infty}.
        \label{eq:l1_lower_bound}
    \end{equation}
    For any $x^{\star}$ such that $|g(x^{\star})| = \|g\|_{\infty}$, the lower bound in Eq.~\eqref{eq:l1_lower_bound} is saturated if and only if there exists a common phase $\phi \in \mathbb{R}$ such that
    \begin{equation}
         c_{k}e^{ikx^\star} = |c_{k}|e^{i\phi}
        \label{eq:phase_condition}
    \end{equation}
\end{lemma}
\begin{proof}
    For any $x\in\Omega_\eta$, the triangle inequality yields
    \begin{equation*}
        |g(x)| = \left| \sum_{k \in \mathbb{Z}} c_k e^{ikx} \right| \leq \sum_{k \in \mathbb{Z}}|c_k|.
    \end{equation*}
    Taking the supremum over $x\in\Omega_\eta$ gives Eq.~\eqref{eq:l1_lower_bound}. Now let $x^{\star}$ satisfy $|g(x^{\star})| = \|g\|_{\infty}$. The bound in Eq.~\eqref{eq:l1_lower_bound} is saturated if and only if the triangle inequality is an equality at $x^\star$:
    \begin{equation*}
        \left| \sum_{k \in \mathbb{Z}} c_k e^{ikx^\star} \right| = \sum_{k \in \mathbb{Z}} |c_k e^{ikx^\star}|.
    \end{equation*}
    By the equality condition for the triangle inequality, this holds if and only if all non-zero terms $c_{k}e^{ikx^{\star}}$ share a common phase $\phi\in\mathbb R$, giving Eq.~\eqref{eq:phase_condition}.
\end{proof}

\begin{lemma}[Convex endpoint reflections are norm-saturating]
\label{lem:convex_endpoint_reflection_saturation}
    Fix $\eta=2$, let $g:\Omega_2\to\mathbb C$ and let $x^\star=q\pi/2$ with $q\in\{-1,1\}$. Suppose there exists $\phi\in\mathbb R$ such that
    \begin{equation}
        h(y)=e^{-i\phi}g(x^\star-qy),\qquad 0\leq y\leq\pi,
    \end{equation}
    is real-valued, convex, non-increasing, has finite one-sided endpoint derivatives, and satisfies $\int_0^\pi h(y)\,{\rm d}y \geq 0$. Define the $2\pi$-periodic endpoint reflection
    \begin{equation}
        F_2(x^\star+qy)=e^{i\phi}h(|y|),\qquad -\pi\leq y\leq\pi.
    \end{equation}
    Then $F_2 = \sum_{k\in\mathbb Z}c_ke^{ikx}$ is a norm-saturating extension of $g$ whose coefficients satisfy
    \begin{equation}
        \sum_{k\in\mathbb Z}|c_k|=|g(x^\star)|=\|g\|_\infty.
        \label{eq:convex_endpoint_reflection_saturation}
    \end{equation}
\end{lemma}
\begin{proof}
    Since $F_2(x^\star+qy)=e^{i\phi}h(|y|)$ is even in $y$, it has a cosine expansion
    \begin{equation*}
        F_2(x^\star+qy) = e^{i\phi}\left(b_0+\sum_{k=1}^\infty b_k\cos(ky)\right),
    \end{equation*}
    where
    \begin{equation*}
        b_0=\frac{1}{\pi}\int_0^\pi h(y)\,{\rm d}y,\qquad
        b_k=\frac{2}{\pi}\int_0^\pi h(y)\cos(ky)\,{\rm d}y.
    \end{equation*}
    The assumption $\int_0^\pi h(y)\,{\rm d}y\geq0$ gives $b_0\geq0$. For $k\geq1$, integration by parts in the Riemann-Stieltjes sense gives
    \begin{equation*}
        b_k = \frac{2}{\pi k^2} \left( h_-'\!(\pi)\left[(-1)^k-1\right] + 2\int_0^\pi \sin^2\!\left[\frac{ky}{2}\right]dh'(y)
        \right)\!.
    \end{equation*}
    Since $h$ is non-increasing, $h_-'\!(\pi)\leq0$, and since $h$ is convex, $d h'(y)\geq0$. Thus $b_k\geq0$ for all $k\geq0$, so
    \begin{equation*}
        |F_2(x^\star+qy)| \leq b_0+\sum_{k=1}^\infty b_k = |F_2(x^\star)| = |g(x^\star)|.
    \end{equation*}
    As $F_2=g$ on $\Omega_2$, this shows that $x^\star$ is a maximizer of $|g|$ on $\Omega_2$.

    Writing $F_2(x)=\sum_{k\in\mathbb Z}c_ke^{ikx}$ and expanding the cosines gives
    \begin{equation*}
        c_0=e^{i\phi}b_0,\qquad
        c_k=e^{i\phi}\frac{b_{|k|}}{2}e^{-ikx^\star},\quad k\neq0,
    \end{equation*}
    so
    \begin{equation*}
        \sum_{k\in\mathbb Z}|c_k| = b_0+\sum_{k=1}^\infty b_k = |g(x^\star)| = \|g\|_\infty,
    \end{equation*}
    as required. 
\end{proof}

\begin{proof}[Proof of Theorem~\ref{thm:optimal_lcu_subnormalization}]
    Using the rescaling in Eqs.~\eqref{eq:G_tau_definition} and \eqref{eq:g_definition} with $\eta=2$, we have $f(\mathcal H)=g(\mathcal G)$ and $f_m(\mathcal H)=g_m(\mathcal G)$. Let $x^\star\in\{-\pi/2,\pi/2\}$ denote the endpoint at which $|g|$ is maximized. Since the spectral interval of $\mathcal H$ is mapped onto $\Omega_2$, we have $x^\star\in\operatorname{spec}(\mathcal G)$. By Lemma~\ref{lem:convex_endpoint_reflection_saturation}, $\alpha_\infty=|g(x^\star)|=\|g\|_\infty$. Since $\operatorname{spec}(\mathcal G)\subseteq\Omega_2$, the spectral theorem gives $\alpha_\infty=\|g(\mathcal G)\|=\|f(\mathcal H)\|$. 
    
    For truncation order $m$, $g(x)-g_m(x)=\sum_{|k|>m}c_ke^{ikx}$. The triangle inequality gives
    \begin{equation*}
        |g(x)-g_m(x)|\leq \sum_{|k|>m}|c_k|=\alpha_\infty-\alpha_m,
        \qquad x\in\Omega_2.
    \end{equation*}
    By the phase alignment from Lemma~\ref{lem:norm_saturation_criterion}, this bound is attained at $x^\star$. Therefore, using the spectral error definition in Eq.~\eqref{eq:spectral_error},
    \begin{equation*}
        \epsilon = \|g(\mathcal G)-g_m(\mathcal G)\| = \sum_{|k|>m}|c_k| = \alpha_\infty-\alpha_m .
    \end{equation*}
    Hence the degree-$m$ reflected-extension Fourier LCU has subnormalization $\alpha=\alpha_m=\|f(\mathcal H)\|-\epsilon$, which is optimal.
\end{proof}

Taking $m\to\infty$ recovers the asymptotic statement
\begin{equation}
    \alpha_\infty^{\rm refl}=\|f(\mathcal H)\|.
\end{equation}
This construction is most useful for functions satisfying the convex endpoint-reflection condition, including $f(x)=x$, $e^x$, and $x^{-1}$. Concave functions such as $\sqrt{x}$ or $\log x$ still achieve a subnormalization that is bounded independently of the error, but this is not necessarily optimal. These functions are better treated using the arcsine-Taylor or regularized constructions in the following sections.

\section{Exponential convergence with bounded subnormalization}
\label{sec:exponential_convergence_with_bounded_subnormalization}

The reflected extension gives optimal subnormalization, but its convergence is algebraic because the continuation is not differentiable at $x=\pm\pi/2$. Scaling the spectrum to be contained compactly within $(-\pi/2,\pi/2)$ improves the algebraic convergence by one order, but the asymptotic subnormalization becomes suboptimal. Algebraic convergence is a consequence of prescribing an exact periodic extension and approximating it globally with a Fourier series. In this section, we use a coefficient sequence that prioritizes the function fit on $(-\pi/2,\pi/2)$ instead of the full $\mathbb{T}$. This achieves exponential convergence on $\Omega_\eta$ with a subnormalization that remains bounded independently of the error. As with the previous section, we first give the construction for the identity function, and then formalize and generalize to other $f$. 

\subsection{Arcsine-Taylor identity function approximation}
\label{sec:arcsine_taylor_identity_function_decomposition}

The regular triangular wave from the previous section may also be written as $f(x)=\arcsin(\sin x)$, which equals the identity function on $\Omega_\eta$. Approximating $\arcsin z$ by its Taylor expansion, then substituting $z=\sin x$, gives
\begin{equation}
    x = \sum_{k=0}^{T} \frac{\binom{2k}{k}}{4^k(2k+1)}\sin^{2k+1}x + O\!\left(\! T^{-3/2}\sin^{2T+1}\left[\frac{\pi}{\eta}\right] \right)
    \label{eq:arcsin_taylor}
\end{equation}
on $\Omega_\eta$, with Taylor truncation order $T$. The truncation error in Eq.~\eqref{eq:arcsin_taylor} is strictly dependent on taking $\eta>2$; using $\eta=2$ gives weak algebraic convergence of only $O(T^{-1/2})$. Each term $\sin^{2k+1}x$ is a finite sine series,
\begin{equation}
    \sin^{2k+1}x = \frac{1}{4^k}\sum_{j=0}^k(-1)^j\binom{2k+1}{k-j}\sin([2j+1]x),
    \label{eq:sine_power_expansion}
\end{equation}
so Eq.~\eqref{eq:arcsin_taylor} can be written as
\begin{equation}
    x = \sum_{k=0}^{T} c_{2k+1}^{(T)} \sin([2k+1]x) + O\!\left( T^{-3/2}\sin^{2T+1}\!\left[\frac{\pi}{\eta}\right] \right),
    \label{eq:arcsin_sine_series}
\end{equation}
with coefficients 
\begin{equation}
    c_{2k+1}^{(T)} = (-1)^k \sum_{j=k}^{T} \frac{\binom{2j}j\binom{2j+1}{j-k}}{16^j(2j+1)}, \qquad 0\leq k\leq T.
    \label{eq:arcsin_lcu_coefficients}
\end{equation}
The even sine modes vanish, and the coefficients $c_{2k+1}^{(T)}$ depend on the truncation order $T$, unlike in a conventional Fourier series.

For consistency with the previous Fourier decompositions, we write the largest mode as $m=2T+1 = \Theta(T)$, so Eq.~\eqref{eq:arcsin_taylor} gives the maximum scalar error on $\Omega_\eta$ as
\begin{equation}
    \epsilon_m^{\rm asin}= O\!\left(m^{-3/2}\sin^m(\pi/\eta)\right),
    \label{eq:arcsin_error_m}
\end{equation}
for fixed $\eta>2$. The approximation therefore converges exponentially in the number of Fourier modes, with convergence factor $\rho_\eta^{\rm asin} =  \sin(\pi/\eta)\in (0,1)$.

Using the scaling notation in Eq.~\eqref{eq:G_tau_definition} and Eq.~\eqref{eq:g_definition}, $\mathcal{H}=\mu I+(\eta\delta/\pi)\mathcal G$, where $\operatorname{spec}(\mathcal G)\subseteq\Omega_\eta$. The constant part $\mu I$ is represented exactly, and $\mathcal G$ by the approximation in Eq.~\eqref{eq:arcsin_sine_series} gives
\begin{align}
    \mathcal{H} &= \mu I+\frac{\eta\delta}{2\pi i}\sum_{k=0}^{T}c_{2k+1}^{(T)}\left(e^{i(2k+1)\mathcal{G}}-e^{-i(2k+1)\mathcal{G}}\right) \nonumber\\
    &+ O\!\left(\eta\delta m^{-3/2}\sin^m(\pi/\eta)\right), \qquad m=2T+1.
    \label{eq:arcsin_lcu}
\end{align}
The corresponding LCU subnormalization is
\begin{equation}
    \alpha_m^{\mathrm{asin}} = |\mu|+\frac{\eta\delta}{\pi}\sum_{k=0}^{T}|c_{2k+1}^{(T)}| = |\mu|+\frac{\eta\delta}{\pi}\sum_{k=0}^{T}\frac{\binom{2k}{k}}{4^k(2k+1)}.
    \label{eq:arcsin_subnormalization_definition}
\end{equation}
Equation~\eqref{eq:arcsin_subnormalization_definition} holds because $c_{2k+1}^{(T)}$ has sign $(-1)^k$, while each power $\sin^{2n+1}x$ has sine-coefficient $\ell^1$-norm equal to one. Taking $m\to\infty$, the sum converges to $\pi/2$, so the coefficient $\ell^1$-norm is bounded uniformly in $m$. The asymptotic subnormalization and its ratio to the optimal value $\alpha_\infty^{\rm refl}$ achieved by the approach in the previous section are
\begin{equation}
    \alpha_\infty^{\mathrm{asin}} = |\mu|+\frac{\eta}{2}\delta, \qquad
    \frac{\alpha_\infty^{\rm asin}}{\alpha_\infty^{\rm refl}} = 1 + \frac{(\eta-2)\delta}{2\|\mathcal H\|},
    \label{eq:action_arcsin_subnormalization_bound}
\end{equation}
where $\|\mathcal H\| = |\mu| + \delta$ from Eq.~\eqref{eq:mu_delta_definition}. The cost of exponential convergence is an $\eta$ dependence on the subnormalization beyond the algebraically optimal value, which is a compromise with the rate of exponential convergence in Eq.~\eqref{eq:arcsin_error_m}. Equation~\eqref{eq:action_arcsin_subnormalization_bound} also reveals that the subnormalization approaches the optimal value with small $\delta$ corresponding to a narrow spectrum, even if $\|\mathcal H\|$ is large.

\subsection{Application to other functions}

The identity construction is a special case of a more general mechanism of approximating $g(x)$ by a Taylor expansion of $g(\arcsin z)$ after the change of variables $z=\sin x$. 

\begin{lemma}[Sine-power Fourier coefficient bounds]
    \label{lem:taylor_sine_bounds}
    Let $g(\arcsin z)=\sum_{k=0}^{\infty}d_k z^k$, where $\{d_{k}\}_{k \in \mathbb N_0} \in \ell^{1}$. Define the truncated arcsine-Taylor approximation of $g$ by a change of variables $z=\sin x$:
    \begin{equation}
        g_m(x)=\sum_{k=0}^md_k\sin^k x = \sum_{k=-m}^mc_k^{(m)}e^{ikx}.
        \label{eq:taylor_sine_approximation}
    \end{equation}
    Then, $g_m$ has a Fourier coefficient $\ell^1$-norm 
    \begin{equation}
        \sum_{k=-m}^m\left|c_k^{(m)}\right| \leq \sum_{k=0}^m|d_k|.
        \label{eq:taylor_sine_l1_bound}
    \end{equation}
\end{lemma}
\begin{proof}
    Since $\sin x=(e^{ix}-e^{-ix})/(2i)$, for each $k\ge 0$
    \begin{equation*}
        \sin^k x=\frac{1}{(2i)^k}\sum_{\ell=0}^{k}(-1)^\ell\binom{k}{\ell}e^{i(k-2\ell)x}.
    \end{equation*}
    Hence $\sin^k x$ is a trigonometric polynomial of degree at most $k$, and the sum of the absolute values of its Fourier coefficients is $2^{-k}\sum_{\ell=0}^{k}\binom{k}{\ell}=1$. Therefore $g_m$ has degree at most $m$, proving Eq.~\eqref{eq:taylor_sine_approximation}. The triangle inequality therefore gives
    \begin{equation*}
        \sum_{k=-m}^m |c_k^{(m)}| \leq \sum_{k=0}^m |d_k| \left[2^{-k}\sum_{\ell=0}^{k}\binom{k}{\ell}\right] = \sum_{k=0}^m |d_k|,
    \end{equation*}
    proving Eq.~\eqref{eq:taylor_sine_l1_bound}.
\end{proof}

\begin{theorem}[Exponential convergence and error-independent subnormalization]
\label{thm:taylor_sine_lcu}
    Suppose that $g(\arcsin z)=\sum_{k=0}^{\infty}d_kz^k$ with $\{d_k\}_{k\geq0}\in\ell^1$. Fix $\eta>2$, let $\mathcal G=\mathcal G^\dagger$ with $\operatorname{spec}(\mathcal G)\subseteq\Omega_\eta$, and let $\rho_\eta=\sin(\pi/\eta)$. Arcsine-Taylor gives an LCU approximation $g_m(\mathcal G)$ of $g(\mathcal G)$ with exponential convergence \begin{equation}
        \|g(\mathcal G)-g_m(\mathcal G)\| \leq \sum_{k=m+1}^{\infty}|d_k|\rho_\eta^k \leq \rho_\eta^{m+1}\sum_{k=0}^\infty |d_k|.
        \label{eq:arcsine_taylor_convergence}
    \end{equation}
    The subnormalization is bounded independently of the error by
    \begin{equation}
        \alpha_m = \sum_{k=-m}^m|c_k^{(m)}| \leq \sum_{k=0}^m|d_k|.
        \label{eq:arcsine_taylor_subnormalization_definition}
    \end{equation}
\end{theorem}
\begin{proof}
    Applying the functional calculus gives $g_m(\mathcal G)=\sum_{|k|\leq m}c_k^{(m)}e^{ik\mathcal G}$. Since $\mathcal G = \mathcal G^\dagger$, each $e^{ik\mathcal G}$ is unitary, so this is an LCU decomposition. For $x\in\Omega_\eta$, we have $|x|<\pi/2$, so $\arcsin(\sin x)=x$. Hence the truncation error is determined by the coefficient tail, $g(x)-g_m(x) = \sum_{k=m+1}^{\infty}d_k\sin^k x$. Since $|\sin x|\leq \rho_\eta$ on $\Omega_\eta$,
    \begin{equation*}
        |g(x)-g_m(x)|\leq \sum_{k=m+1}^{\infty}|d_k|\rho_\eta^k.
    \end{equation*}
    Taking the supremum over $x\in\Omega_\eta$ and applying the spectral theorem gives Eq.~\eqref{eq:arcsine_taylor_convergence}. Since $\eta>2$, $\rho_\eta \in (0,1)$, Eq.~\eqref{eq:arcsine_taylor_convergence} is exponentially convergent. The subnormalization bound in Eq.~\eqref{eq:arcsine_taylor_subnormalization_definition} follows from Lemma~\ref{lem:taylor_sine_bounds}.
\end{proof}

\begin{corollary}[Endpoint-saturated subnormalization]
    \label{cor:endpoint_saturated_asymptotic_subnormalization}
     Under the assumptions of Theorem~\ref{thm:taylor_sine_lcu}, let $\mu$ and $\delta$ be the spectral center and half-width from Eq.~\eqref{eq:mu_delta_definition}. Suppose $f$ satisfies the convex endpoint-maximizing condition of Lemma~\ref{lem:convex_endpoint_reflection_saturation}. Then, the asymptotic subnormalization in Theorem~\ref{thm:taylor_sine_lcu} is 
    \begin{equation}
        \alpha_\infty^{\rm asin} = \left| f\!\left(\mu+q\frac{\eta\delta}{2}\right) \right|,
        \label{eq:arcsine_taylor_subnormalization}
    \end{equation}
    where $q\in\{-1,1\}$ gives the saturating endpoint. 
\end{corollary}
\begin{proof}
    $g(\arcsin\sin x)$ is a reflected extension about $x\pm\pi/2$, so by Theorem~\ref{thm:optimal_lcu_subnormalization}, the convex endpoint-reflected extension is norm-saturating. Therefore,
    \begin{equation*}
        \alpha_\infty = \sum_{k\in\mathbb Z}|c_k^{(\infty)}| = |g(x^\star)|.
    \end{equation*} 
    Taking $x^\star=q\pi/2$ and defining $g(x)$ from Eq.~\eqref{eq:g_definition} gives Eq.~\eqref{eq:arcsine_taylor_subnormalization}.
\end{proof}

Theorem~\ref{thm:taylor_sine_lcu} shows that a sufficient condition for error-independent subnormalization with exponential convergence is an absolutely convergent Taylor series $\{d_{k}\}_{k \in \mathbb{N}_0} \in \ell^{1}$. In the rescaled matrix-function setting, we take $\mathcal G=\tau(\mathcal H-\mu I)$ and $g(x)=f(\mu+x/\tau)$, so that $f(\mathcal H)=g(\mathcal G)$. Therefore, 
\begin{equation}
    g(\arcsin z) = f\left(\mu + \frac{\eta\delta}{\pi}\arcsin z\right)
    \label{eq:h_function}
\end{equation}
has at most square-root boundary singularities inherited from $\arcsin z$ at $z=\pm1$, which give $d_k=O(k^{-3/2})$. Hence, $\{d_k\}_{k\in\mathbb N_0}\in\ell^1$ when $f$ admits an analytic continuation to a complex neighborhood of the compact set
\begin{align}
    \left\{\mu + \frac{\eta \delta}{\pi} \arcsin z: |z| \leq 1 \right\}\! &\subseteq  \Omega_\eta^{\rm asin}
    \!+ i\frac{\eta\delta}{\pi}\operatorname{asinh}(1)\left[-1,1\right]\!,
    \label{eq:set_for_analytic_f} 
\end{align}
where 
\begin{equation}
    \Omega_\eta^{\rm asin} = \left[\mu-\frac{\eta\delta}{2}, \mu+\frac{\eta\delta}{2}\right].
    \label{eq:omega_eta_asin}
\end{equation}
When $f$ is entire, such as for $f(x)=x$ or $f(x)=e^x$, this holds for any $\eta>2$. When $f$ is non-entire, we may often be able to choose $\eta>2$ so that Eq.~\eqref{eq:set_for_analytic_f} remains inside a complex domain of analyticity of $f$. For example, the functions $f(x) = x^{-1}$, $\log x$ and $\sqrt{x}$ are holomorphic for the open right half-plane. Therefore, $\Omega_\eta^{\rm asin}$ must be entirely positive, so $2<\eta < 2\mu/\delta$ for $\mu/\delta = (\kappa+1)/(\kappa-1)$. Choosing a large $\eta$ increases the rate of exponential convergence, but requires fitting closer to the singularity or branch point. 
    
\section{Sobolev-regularized extensions}
\label{sec:sobolev_regularised_extensions}

The approaches thus far approximate the target function on $\Omega_\eta$ by prescribing a function extension on the full period $\mathbb T$. In this section, we no longer specify the full periodic function, but find the coefficients as the solution of a regularized approximation problem by fitting on $\Omega_\eta$ only. The simplest method for computing Fourier-extension coefficients is by least squares, which is analogous to the $L^2$ convergence of standard Fourier series~\cite{Adcock2014numerical, Webb2020}. While this approximation converges exponentially in $m$ on the fitted interval, it can become highly oscillatory outside of it, with Fourier coefficients that do not necessarily decay~\cite{Huybrechs2010}. Least squares is therefore a poor coefficient selection strategy in general, and particularly when minimizing the coefficient $\ell^1$-norm alongside the approximation error is the priority. We instead develop a regularized coefficient selection strategy that achieves near-exponential convergence, while also providing a mechanism to bound the subnormalization at a target error tolerance towards near-optimal values. The approximation domain may be chosen more flexibly than the previous approaches, using a finite union of closed intervals $\Omega$ where $\operatorname{spec}(\mathcal G)\subseteq\Omega$. This is essential for cases where $\mathcal H$ has eigenvalues that lie on both sides of a pole or branch point, such as indefinite matrix inversion. To aid the comparison to the coefficient-selection strategies presented in Sections~\ref{sec:optimal_subnormalization_on_extended_domains} and \ref{sec:exponential_convergence_with_bounded_subnormalization}, we derive the convergence on $\Omega_\eta$ where applicable, and give examples for unions of spectral intervals in Section~\ref{sec:application_to_various_functions}.

In a standard, non-extended Fourier series, the degree-$m$ approximation space is
\begin{equation}
    \mathcal{F}_m= \mathrm{span}\{e^{-imx},\dots,e^{imx}\} \subset L^{2}(\mathbb{T}),
\end{equation}
whose Fourier modes form an orthogonal basis for $\mathcal F_m$. This yields a unique mapping from functions to coefficients. For a Fourier extension, the Fourier basis is defined over the full period $\mathbb{T}$, and the approximation domain is restricted to $\Omega_\eta$ defined in Eq.~\eqref{eq:omega_eta}. When restricted to $\Omega_\eta$, the infinite system is redundant, and is therefore a frame rather than an orthogonal basis. For a fixed truncation $\mathcal F_m$, the associated least-squares map becomes increasingly ill-conditioned with $m$, so large changes in the coefficient vector can have very small effects on the fit over $\Omega_\eta$. Regularizing the coefficients (i.e., imposing a constraint on their norm) provides a way to select favorable coefficients with low subnormalization $\alpha$ for any given error $\epsilon_m$. Since $\alpha$ is a function of the $\ell^{1}$-norm of the Fourier coefficients, the natural choice of loss function for such a regularization process is
\begin{equation}
    J(\mathbf c;m,\gamma,\eta)=\underbrace{\|g-g_m\|_{L^\infty(\Omega_\eta)}}_{\epsilon_m}+\gamma\underbrace{\|\mathbf c\|_1}_{\alpha},
    \label{eq:lasso}
\end{equation}
where $\gamma>0$ is a tunable hyperparameter controlling the error-subnormalization trade-off. However, both of these norms ($L^{\infty}$ and $\ell^{1}$) are challenging to optimize for to high accuracy as they are non-smooth. The $L^{2}$ and $\ell^{2}$-norms are smooth and therefore common in many applications, but do not guarantee control over $L^{\infty}$ and $\ell^{1}$.

Instead, we look to Sobolev norms~\cite{Adams2003}, where a combination of the function and its derivatives (or fractional derivatives, as below) is penalized.
\begin{definition}[Weighted fractional Sobolev norm]
    \label{def:weighted_sobolev_norm}
    Let $0<r\leq 1$ and $w>1/2$. For a periodic function $f$ on $\mathbb T$ with Fourier coefficients $\hat f_k$, the $r$-weighted fractional Sobolev norm is given by
    \begin{equation}
        \|f\|_{H^w_r(\mathbb T)}^2 = 2\pi\sum_{k\in\mathbb Z}\left(1+(r|k|)^{2w}\right)|\hat f_k|^2.
        \label{eq:scaled_Hw_regularizer}
    \end{equation}
\end{definition}
Using $r=1$ gives the standard periodic fractional Sobolev norm. The case $w=1$ recovers the weighted $H^1_r$ norm, which, using Parseval's identity, can be written as $\|f\|_{H_r^1(\mathbb T)}^2 = \|f\|_{L^2(\mathbb T)}^2 + r^2\|f'\|_{L^2(\mathbb T)}^2$. Using $w>1/2$ gives a fractional regularizer that is still strong enough to control the coefficient $\ell^1$-norm. The scale $r$ is chosen according to the shortest length scale that must be resolved by the fitted function, which allows the Fourier-coefficient $\ell^1$-norm to scale optimally. By using the $H^1$-norm for fitting and the $H^w_r$-norm for coefficient regularization, we can converge in $L^\infty$ and control the $\ell^1$-norm of the Fourier coefficients.

\begin{lemma}[$\ell^{1}$ boundedness via fractional Sobolev norm]\label{thm:l1h1}
    Let $1/2<w\leq 1$ and $0<r\leq 1$. For a given Fourier series $g_m$ that is periodic on $\mathbb{T}$, the $\ell^{1}$-norm of its coefficient vector $\mathbf{c}$ is bounded by the $H^{w}_{r}$-norm of $g_m$:
    \begin{equation}
        \|\mathbf{c}\|_{1} \leq C_w r^{-1/2}\|g_m\|_{H^{w}_{r}(\mathbb{T})}.
        \label{eq:l1_scaled_Hw_bound}
    \end{equation}
\end{lemma}
\begin{proof}
    From Definition~\ref{def:weighted_sobolev_norm},
    \begin{equation*}
        \|g_m\|_{H^{w}_{r}(\mathbb{T})}^{2} = 2\pi\sum_{k= -m}^m\left(1+(r|k|)^{2w}\right)|c_{k}|^{2}.
    \end{equation*}
    Applying Cauchy-Schwarz to the $\ell^1$-norm of the Fourier coefficients gives
    \begin{equation*}
        \|\mathbf{c}\|_{1} \leq \!\left(\sum_{k=-m}^m\!\frac{1}{1+(r|k|)^{2w}}\!\right)^{\!\!\frac{1}{2}} \!\!\!\left(\sum_{k=-m}^m\!\!(1+(r|k|)^{2w})|c_k|^2\!\right)^{\!\!\frac{1}{2}}\!\!.
    \end{equation*}
    Since $w>1/2$, the integral $I_w=\int_0^\infty(1+t^{2w})^{-1}{\rm d}t$ is finite. The summand is decreasing in $|k|$, so for $0<r\leq 1$
    \begin{equation*}
        \sum_{k\in\mathbb Z}\frac{1}{1+(r|k|)^{2w}} \leq 1+2r^{-1}I_w \leq C'_w r^{-1}.
    \end{equation*}
    This gives Eq.~\eqref{eq:l1_scaled_Hw_bound} with $C_w=(C'_w/[2\pi])^{1/2}$.
\end{proof}

Equation~\eqref{eq:l1_scaled_Hw_bound} satisfies $\|\mathbf c\|_1=O(r^{-1/2}\|g_m\|_{H^w_r(\mathbb T)})$. Therefore, if the scaled regularizer satisfies
\begin{equation}
    \|g_m\|_{H^w_r(\mathbb T)}=O(\|g\|_{L^\infty(\Omega)}r^{1/2}),
    \label{eq:sobolev_uniform_bound}
\end{equation}
the resulting coefficient $\ell^1$-norm has the optimal scale $\|\mathbf c\|_1=O(\|g\|_{L^\infty(\Omega)})$. Spectral tightness and an eigenvalue located at the function maximum $\|g\|_{L^\infty(\Omega)}=\max_{\lambda\in\operatorname{spec}(\mathcal G)}|g(\lambda)|=\|g(\mathcal G)\|$ then gives optimal subnormalization dependence $\alpha=O(\|f(\mathcal H)\|)$, since $f(\mathcal H)=g(\mathcal G)$. The role of $r$ is therefore to counteract the Sobolev penalty at the shortest length scale, promoting the scale balance in Eq.~\eqref{eq:sobolev_uniform_bound}. This is most easily seen for $w=1$, where $\|g_m\|_{H^1_r(\mathbb T)}^2=\|g_m\|_{L^2(\mathbb T)}^2+r^2\|g_m'\|_{L^2(\mathbb T)}^2$. If a function varies on a length scale $r$, the derivative norm may be larger than the function norm by a factor $O(r^{-1})$, so the factor $r^2$ places the derivative contribution on the same scale as the function contribution. We take $w=1$ whenever the natural scale $r$ gives Eq.~\eqref{eq:sobolev_uniform_bound}. When the $H^1_r$ penalty is too strong, we may instead take a fixed $1/2<w<1$ to weaken the derivative penalty while retaining control of the coefficient $\ell^1$-norm, although at a cost of increasing $C_w$ in Eq.~\eqref{eq:l1_scaled_Hw_bound}. We motivate choices of $r$ for several functions in Section~\ref{sec:application_to_various_functions}, and prove Theorem~\ref{thm:sobolev_complexity} in Section~\ref{sec:resource_requirements}.

Instead of the loss function in Eq.~\eqref{eq:lasso}, we define $\tilde{g}_m$ using Sobolev norms as 
\begin{equation}\label{eq:fn}
    \tilde{g}_m\in \underset{h \in \mathcal{F}_m}{\arg\min}\!\left\{\|g-h\|^{2}_{H^{1}(\Omega_\eta)} + \gamma\|h\|_{H^{w}_{r}(\mathbb{T})}^{2}\right\}.
\end{equation}

\begin{lemma}[Regularized minimizer bounds]\label{thm:minimizer}
Any $\tilde{g}_m$ given by Eq.~\eqref{eq:fn} satisfies
\begin{equation}
    \|g-\tilde{g}_m\|_{H^{1}(\Omega_\eta)}^{2} \leq \underset{h \in \mathcal{F}_m}{\inf}\left(\|g-h\|^{2}_{H^{1}(\Omega_\eta)} + \gamma\|h\|_{H^{w}_{r}(\mathbb{T})}^{2}\right)
    \label{eq:regularized_minimizer_error_bound}
\end{equation}
and
\begin{equation}
    \|\tilde{g}_m\|_{H^{w}_{r}(\mathbb{T})}^{2} \leq \underset{h \in \mathcal{F}_m}{\inf}\left(\frac{\|g-h\|^{2}_{H^{1}(\Omega_{\eta})}}{\gamma} + \|h\|_{H^{w}_{r}(\mathbb{T})}^{2}\right).
    \label{eq:regularized_minimizer_hw_bound}
\end{equation}
\end{lemma}
\begin{proof}
    Since $\tilde{g}_m$ is a minimizer, we have
    \begin{equation*}
    \begin{split}
        &\|g-\tilde{g}_m\|^{2}_{H^{1}(\Omega_\eta)} + \gamma\|\tilde{g}_m\|_{H_r^{w}(\mathbb{T})}^{2} \\ &\leq \|g-h\|^{2}_{H^{1}(\Omega_\eta)} + \gamma\|h\|_{H_r^{w}(\mathbb{T})}^{2},\quad  \forall h \in \mathcal{F}_m.
        \end{split}
    \end{equation*}
    Since both terms on the left-hand side are nonnegative, either term individually satisfies the inequality.
\end{proof}

\begin{theorem}[Existence of $H^1$ approximation]\label{thm:unreg_superalg}
    Let $k>1$ and $g\in H^k(\Omega_\eta)$. Then, for each $m\in\mathbb N$, there exists an $h\in\mathcal F_m$ such that
    \begin{align}
        \|g-h\|_{H^1(\Omega_\eta)} &\leq C_{k,\eta}m^{1-k}\|g\|_{H^k(\Omega_\eta)}, \\
        \|h\|_{H^k(\mathbb T)} &\leq C_{k,\eta}\|g\|_{H^k(\Omega_\eta)}.
    \end{align}
\end{theorem}

\begin{proof}
    We follow the proof given in \cite[Proposition 17]{Adcock2019}. There is an extension $u \in H^{k}(\mathbb{T})$ of $g$ with $\| u\|_{H^k(\mathbb{T})} \leq C_{k,\eta} \| g \|_{H^k(\Omega_\eta)}$. Let $c_j$ be the Fourier coefficients of $u$ on $\mathbb{T}$ and define $h = \sum^m_{j=-m} c_j e^{ijx} \in \mathcal{F}_m$. 

    Then, by Definition~\ref{def:weighted_sobolev_norm}
    \begin{align*}
        \|h\|_{H^k(\mathbb T)}^2 &= 2\pi\sum_{|j|\leq m}|c_j|^2(1+j^2)^k\\
        &\leq2\pi\sum_{j\in\mathbb Z}|c_j|^2(1+j^2)^k\\
        &=\|u\|_{H^k(\mathbb T)}^2\leq C_{k,\eta}^2\|g\|_{H^k(\Omega_\eta)}^2.
    \end{align*}
    Also,
    \begin{align*}
        \|g-h\|_{H^1(\Omega_\eta)}^2 &\leq\|u-h\|_{H^1(\mathbb T)}^2=2\pi\sum_{|j|>m}|c_j|^2(1+j^2)\\
        &\leq(1+m^2)^{1-k}2\pi\sum_{|j|>m}|c_j|^2(1+j^2)^k\\
        &\leq(1+m^2)^{1-k}\|u\|_{H^k(\mathbb T)}^2,
    \end{align*}
    from which we deduce that
    \begin{align*}
        \| g-h \|_{H^1(\Omega_\eta)} &\leq C_{k,\eta} (1+m^2)^{\frac{1-k}{2}} \|g \|_{H^k(\Omega_\eta)}\\ 
        &\leq C_{k,\eta} m^{1-k} \|g \|_{H^k(\Omega_\eta)},
    \end{align*}
    as required.
\end{proof}

We can use this theorem to demonstrate convergence properties of the regularized problem. We define a Sobolev embedding constant
\begin{equation}
    C_\eta = \sqrt{\frac{\eta}{2\pi}+\frac{2\pi}{\eta}},
    \label{eq:sobolev_embedding_constant}
\end{equation}
so that $\|u\|_{L^\infty(\Omega_\eta)} \leq C_\eta\|u\|_{H^1(\Omega_\eta)}$ for $u\in H^1(\Omega_\eta)$.

\begin{lemma}[Superalgebraic convergence and coefficient bounds for Sobolev-regularized extensions]
    Suppose $g \in H^{k}(\Omega_{\eta})$ for some $k > 1$, let $1/2<w\leq 1$, let $\tilde{g}_m$ be given by Eq.~\eqref{eq:fn}, and let $C_w$ be from Lemma~\ref{thm:l1h1}. Then
    \begin{equation}
        \begin{split}
            \|g-\tilde{g}_m\|_{L^\infty(\Omega_\eta)} &\leq C_\eta\|g-\tilde{g}_m\|_{H^1(\Omega_\eta)} \\
            &\leq C_\eta C_{k,\eta} \left(m^{1-k}+\gamma^{1/2}\right) \|g\|_{H^k(\Omega_\eta)}.
        \end{split}
    \end{equation}
    and
    \begin{align}
        \|\mathbf{c}\|_{1} \leq C_w C_{k,\eta} r^{-1/2}\left(\frac{m^{1-k}}{\gamma^{1/2}} + 1\right)\|g\|_{H^{k}(\Omega_{\eta})},
    \end{align}
\end{lemma}

\begin{proof}
    Let $h$ be as given in Theorem~\ref{thm:unreg_superalg}. Then
    \begin{align*}
        \|g-h\|_{H^{1}(\Omega_\eta)} \leq C_{k,\eta} m^{1-k}&\|g\|_{H^{k}(\Omega_\eta)} \\
        \|h\|_{H^{w}_{r}(\mathbb{T})}^{2} \leq \|h\|_{H^{1}(\mathbb{T})}^{2} \leq \|h\|_{H^{k}(\mathbb{T})}^{2} \leq &\left(C_{k,\eta}\|g\|_{H^{k}(\Omega_\eta)}\right)^{2}. 
    \end{align*}
    We deduce from Lemma~\ref{thm:minimizer} and the triangle inequality that
    \begin{align*}
        \|g-\tilde{g}_m\|_{H^{1}(\Omega_\eta)} &\leq (\|g-h\|_{H^{1}(\Omega_\eta)}^{2} + \gamma\|h\|_{H^{w}_{r}(\mathbb{T})}^{2})^{\frac{1}{2}}\\
        &\leq C_{k,\eta}(m^{2(1-k)} + \gamma)^{\frac{1}{2}}\|g\|_{H^{k}(\Omega_\eta)}\\
        &\leq C_{k,\eta}(m^{1-k} + \gamma^{\frac{1}{2}})\|g\|_{H^{k}(\Omega_\eta)}.
    \end{align*}
    Combining this estimate with the Sobolev embedding $\|u\|_{L^\infty(\Omega_\eta)} \leq C_\eta \|u\|_{H^1(\Omega_\eta)}$ gives the $L^\infty(\Omega_\eta)$ bound.
    
    For the coefficient norm, from Lemma~\ref{thm:l1h1}, we know that
    \begin{equation*}
        \|\mathbf{c}\|_{1} \leq C_w r^{-1/2}\|\tilde{g}_m\|_{H^{w}_{r}(\mathbb{T})}.
    \end{equation*}
    Applying Lemma~\ref{thm:minimizer} and Theorem~\ref{thm:unreg_superalg} again gives the claimed bound.
\end{proof}

Hence, for any $g\in C^\infty(\Omega_\eta)$, we obtain superalgebraic convergence in $m$ down to an $O(\sqrt{\gamma})$ regularization floor. By choosing to minimize the square of the norms, this $H^{1}$-fitting and $H^{w}_{r}$-regularization problem is a form of Tikhonov regularization. If we consider a quadrature rule (e.g., Gauss-Legendre) over the extended domain $\Omega_\eta$ with weights $w_j > 0$ and nodes $x_j$, we can approximate the $H^1$-norm fitting penalty directly on the grid
\begin{equation}
    \|g\|_{H^{1}(\Omega_\eta)}^{2} \approx \sum_{j=1}^{M} w_j \left( |g(x_j)|^{2} + |g'(x_j)|^{2} \right).
\end{equation}
By analytically evaluating the derivative of the target function and the Fourier series, the regularized fitting problem can be written entirely in terms of the grid weights as
\begin{equation}
    \underset{\mathbf{c}'\in\mathbb{C}^{2m+1}}{\min}\|\mathbf{b}_{g}-A_{\tilde{g}_m}\mathbf{c}'\|_{2}^{2} + \|\mathbf{b}_{g'} - A_{\tilde{g}_m'}\mathbf{c}'\|_{2}^{2} + \gamma\|W_{r,w}\mathbf{c}'\|_{2}^{2}
\end{equation}
where
\begin{equation}
    \begin{split}
        (A_{\tilde{g}_m})_{j,k} = \sqrt{w_j}e^{ikx_j}&,(A_{\tilde{g}_m'})_{j,k} = ik\sqrt{w_j}e^{ikx_j}\\ 
        (b_g)_j = \sqrt{w_j}g(x_j)&,(b_{g'})_j = \sqrt{w_j}g'(x_j), \\ 1 \leq j \leq M&, -m \leq k\leq m.
    \end{split}
\end{equation}
Through Parseval's identity, it can be seen that
\begin{equation}
    W_{r,w}=\sqrt{2\pi}\,\operatorname{diag}\!\left(\sqrt{1+(r|k|)^{2w}}\right)_{k=-m}^m,
\end{equation}
and the problem is then equivalent to
\begin{equation}
    \underset{\mathbf{c}'\in\mathbb{C}^{2m+1}}{\min}\left\|
    \begin{bmatrix}
        \mathbf{b}_{g} \\ \mathbf{b}_{g'} \\ \mathbf 0 
    \end{bmatrix} - 
    \begin{bmatrix}
        A_{\tilde{g}_m} \\ A_{\tilde{g}_m'} \\
        \sqrt{\gamma}W_{r,w}
    \end{bmatrix}
    \mathbf{c}'\right\|_{2}^{2}.
\end{equation}
The parameter entering the augmented least-squares system is $\sqrt{\gamma}$, which scales linearly with the approximation error. We therefore tune $\sqrt{\gamma}$ iteratively to attain the prescribed tolerance $\epsilon_{\rm t}$ without representing the squared parameter $\gamma$ explicitly.

As noted earlier, we have demonstrated superalgebraic convergence of the regularized problem, and we can now extend this to near-exponential convergence. First, we need to define the Gevrey class.

\begin{definition}
    Given a compact set $D \subseteq \mathbb{R}$ and $\sigma \geq 1$, the Gevrey class $ G^{\sigma}(D)$ consists of smooth functions $f \in C^{\infty}(D)$ such that there exists a constant $C > 0$ depending on $f,\sigma$ and $D$ only, such that
    \begin{equation}
        \underset{x \in D}{\sup}|f^{(k)}(x)| \leq C^{k+1}(k!)^{\sigma}, \quad k \in \mathbb{N}_{0}
        \label{eq:gev}
    \end{equation}
\end{definition}

Note that $G^{1}(D) = C^{\omega}(D)$, which is the set of analytic functions on $D$. We write $C = C(f,\sigma,D)$ for the smallest constant such that Eq.~\eqref{eq:gev} holds.

\begin{theorem}[Near-exponential convergence of Sobolev-regularized extensions of analytic functions]
    \label{thm:near_exponential_convergence}
    Suppose that $g\in C^\omega(\Omega_{\eta'})$ for some $1<\eta'<\eta$. Consider $\tilde{g}_m$ given by Eq.~\eqref{eq:fn} with $1/2<w\leq1$. Then, for every $\sigma>1$, there exists a periodic extension $h\in G^\sigma(\mathbb T)$ satisfying $h=g$ on $\Omega_\eta$ and constants $C_1,C_2>0$ such that
    \begin{equation}
        \begin{split}
            \|g-\tilde{g}_m\|_{L^\infty(\Omega_\eta)} &\leq C_\eta\|g-\tilde{g}_m\|_{H^1(\Omega_\eta)} \\
            &\leq C_\eta\!\left(\!C_2\exp[-C_1m^{1/\sigma}]+\sqrt{\gamma}\|h\|_{H_r^w\!(\mathbb T)}\!\right)
        \end{split}
        \label{eq:sobolev_near_exponential_error}
    \end{equation}
    and
    \begin{equation}
        \|\mathbf c\|_1\leq C_wr^{-1/2}\left(\frac{C_2\exp[-C_1m^{1/\sigma}]}{\sqrt{\gamma}}+\|h\|_{H_r^w(\mathbb T)}\right).
        \label{eq:sobolev_coefficient_bound}
    \end{equation}
    The constants $C_1$ and $C_2$ are independent of $m$ and $\gamma$, depending only on $g$, $\sigma$, $\eta$, and $\eta'$.
\end{theorem}
\begin{proof}
    For any $\sigma > 1$ there exists a cutoff function $\chi \in G^{\sigma}(\mathbb{R})$ such that $\chi(x) = 1$ for $|x| \leq \pi/\eta$ and $\chi(x) = 0$ for $|x| > \pi/\eta'$. This function can be constructed from the elementary building block
    \begin{equation*}
        u(t) = 
        \begin{cases}
            \exp\left(-t^{1/(1-\sigma)}\right), \quad &t > 0 \\
            0, &t \leq 0.
        \end{cases}
    \end{equation*}
    We first define
    \begin{equation*}
        v(t) = \frac{u(t)}{u(t) + u(1-t)}
    \end{equation*}
    and note that $v(t) = 0$ for $t \leq 0$ and $v(t) = 1$ for $t \geq 1$. Then, we define
    \begin{equation*}
        \chi(x) = v\!\left(\frac{\pi/\eta' - x}{\pi/\eta' - \pi/\eta}\right)\! v\!\left(\frac{\pi/\eta' + x}{\pi/\eta' - \pi/\eta}\right),
    \end{equation*}
    which has the desired properties. Now define the function $h:\mathbb{T}\rightarrow\mathbb{C}$ by
    \begin{equation*}
        h = \begin{cases}
            g(x)\chi(x), \quad &|x| \leq \pi/\eta', \\
            0, &\textrm{otherwise.}            
        \end{cases}
    \end{equation*}
    By construction, $h$ is periodic and Gevrey-$\sigma$, i.e., $h\in G^\sigma(\mathbb T)$, with a Gevrey constant depending only on $g$, $\sigma$, $\eta$, and $\eta'$. Let
    \begin{equation*}
        \widehat h_k=\frac{1}{2\pi}\int_{-\pi}^{\pi}h(x)e^{-ikx}\,{\rm d}x,\qquad h_m(x)=\sum_{k=-m}^m\widehat h_ke^{ikx}.
    \end{equation*}
    Thus, $h_m\in\mathcal F_m$ is the degree-$m$ Fourier truncation of $h$. For $k\neq0$, integrating by parts $\ell$ times and using the Gevrey bound gives
    \begin{equation*}
        |\widehat h_k|\leq C(h,\sigma,\mathbb T)^{\ell+1}(\ell!)^\sigma|k|^{-\ell},\qquad \ell\in\mathbb N_0.
    \end{equation*}
    Stirling's formula and optimization over $\ell$ therefore imply the existence of constants $C_1,C_2>0$ such that
    \begin{equation*}
        |\widehat h_k|\leq C_2\exp[-C_1|k|^{1/\sigma}],\qquad k\in\mathbb Z.
    \end{equation*}
    After increasing $C_2$ and decreasing $C_1$ if necessary, the Fourier tail satisfies
    \begin{equation*}
        \|g-h_m\|_{H^1(\Omega_\eta)}\leq\|h-h_m\|_{H^1(\mathbb T)}\leq C_2\exp[-C_1m^{1/\sigma}].
    \end{equation*}
    Moreover, since $h_m$ is the degree-$m$ Fourier truncation of $h$, Definition~\ref{def:weighted_sobolev_norm} gives $\|h_m\|_{H_r^w(\mathbb T)}\leq\|h\|_{H_r^w(\mathbb T)}$. Applying Eq.~\eqref{eq:regularized_minimizer_error_bound} with $h_m$ as the comparison function gives
    \begin{equation*}
        \begin{split}
            \|g-\tilde{g}_m\|_{H^1(\Omega_\eta)} &\leq\left(\|g-h_m\|_{H^1(\Omega_\eta)}^2+\gamma\|h_m\|_{H_r^w(\mathbb T)}^2\right)^{1/2} \\
            &\leq C_2\exp[-C_1m^{1/\sigma}]+\sqrt{\gamma}\|h\|_{H_r^w(\mathbb T)}.
        \end{split}
    \end{equation*}
    The $L^\infty$ estimate follows from the Sobolev embedding bound defining $C_\eta$. Similarly, Eq.~\eqref{eq:regularized_minimizer_hw_bound} gives
    \begin{equation*}
        \begin{split}
            \|\tilde{g}_m\|_{H_r^w(\mathbb T)} &\leq\left(\frac{\|g-h_m\|_{H^1(\Omega_\eta)}^2}{\gamma}+\|h_m\|_{H_r^w(\mathbb T)}^2\right)^{1/2} \\
            &\leq\frac{C_2\exp[-C_1m^{1/\sigma}]}{\sqrt{\gamma}}+\|h\|_{H_r^w(\mathbb T)}.
        \end{split}
    \end{equation*}
    Lemma~\ref{thm:l1h1} then proves Eq.~\eqref{eq:sobolev_coefficient_bound}.
\end{proof}

The same argument applies when the fitted set is a finite union of disjoint closed intervals $\Omega\subset\mathbb T$, provided that $g$ is analytic on a neighborhood of $\Omega$. We choose a Gevrey cutoff for each connected component and sum the resulting compactly supported extensions. The conclusions of Theorem~\ref{thm:near_exponential_convergence} then hold with $C_\eta$ replaced by the corresponding Sobolev embedding constant $C_\Omega$ of the fitted set. 

For $\|f(\mathcal H)\|>0$, the cost of extending the fitted function periodically is
\begin{equation}
    C_f=\max\left\{1,\,2C_wr^{-1/2}\frac{\|h\|_{H_r^w(\mathbb T)}}{\|f(\mathcal H)\|}\right\},
    \label{eq:Cf_extension_constant}
\end{equation}
which is independent of $m$ and $\epsilon$ and appears in Theorem~\ref{thm:sobolev_complexity}, its proof in Section~\ref{sec:sobolev_resources}, and the following section for various matrix function examples. We call a periodic extension $h\in G^\sigma(\mathbb T)$ scale-adapted at scale $r$ if $h=g$ on $\Omega$ and, for some $C_0\geq1$ independent of the parameters of interest,
\begin{align}
    \|h\|_{H_r^1(\mathbb T)} &\leq C_0 r^{1/2}\|f(\mathcal H)\|, \label{eq:scale_adapted_condition} \\
    \|h^{(\ell)}\|_{L^1(\mathbb T)} &\leq C_0^{\ell+1}(\ell!)^\sigma r^{1-\ell}\|f(\mathcal H)\|,\quad \ell\geq1.
    \nonumber 
\end{align}

\begin{lemma}[Scale-adapted periodic extensions]
    \label{lem:scale_adapted_extension}
    Fix $1/2<w\leq 1$. If $h$ is scale-adapted at scale $r$, then
    \begin{equation}
        r^{-1/2}\|h\|_{H_r^w(\mathbb T)} \leq \sqrt{2}\,C_0\|f(\mathcal H)\|.
        \label{eq:scale_adapted_subnormalization}
    \end{equation}
    and, for $m\geq r^{-1}$,
    \begin{equation}
        \|g-h_m\|_{H^1(\Omega)} = O\!\left( r^{-1/2}\|f(\mathcal H)\| \exp\left[-K(rm)^{1/\sigma}\right]
        \right),
        \label{eq:scale_adapted_extension_tail}
    \end{equation}
    where $h_m$ is the degree-$m$ Fourier truncation of $h$,
    with $K>0$ and the hidden constants depending only on $C_0$
    and $\sigma$.
\end{lemma}
\begin{proof}
    Applying the bound $1+x^{2w}\leq 2(1+x^2)$ for $x\geq0$ and $1/2<w\leq1$ to the Fourier weights in Definition~\ref{def:weighted_sobolev_norm} gives $\|h\|_{H_r^w(\mathbb T)}\leq\sqrt{2}\|h\|_{H_r^1(\mathbb T)}$. Equation~\eqref{eq:scale_adapted_subnormalization} then follows immediately from Eq.~\eqref{eq:scale_adapted_condition}.   
    
    For $k\neq0$, integrating $\widehat{h}_k$ by parts $\ell$ times and applying Eq.~\eqref{eq:scale_adapted_condition} gives $|\widehat h_k|\leq C_0^{\ell+1}(\ell!)^\sigma r\|f(\mathcal H)\|(r|k|)^{-\ell}$. For $r|k|\geq1$, choosing $\ell$ to minimize the upper bound gives $|\widehat h_k|\leq Br\|f(\mathcal H)\|\exp[-K(r|k|)^{1/\sigma}]$ for constants $B,K>0$ depending only on $C_0$ and $\sigma$. Since $m\geq r^{-1}$, this estimate applies to every mode in the Fourier tail. Since $h=g$ on $\Omega$, summing the $H^1$ Fourier tail and decreasing $K$ if necessary gives Eq.~\eqref{eq:scale_adapted_extension_tail}.
\end{proof}

\section{Application to various functions}
\label{sec:application_to_various_functions}

We now give examples for other matrix functions of practical interest: identity $\mathcal H$, exponential $e^{\mathcal H}$, inverse $\mathcal H^{-1}$, and square root $\sqrt{\mathcal H}$. The identity, exponentiation, and inversion examples all achieve the optimal subnormalization scale $\alpha = O(\|f(\mathcal H)\|)$, while the square-root function results in an additional $O(\kappa^{1/4})$ dependence for $w=1$. 

\subsection{Identity (untransformed block encoding)}

The use of the identity function for block encoding the input matrix $\mathcal H$ has already been covered in the context of the optimal subnormalization and arcsine-Taylor approaches in Sections~\ref{sec:triangular_extension_of_the_identity_function} and \ref{sec:arcsine_taylor_identity_function_decomposition}, so here we limit the discussion to the implications for the Sobolev-regularized Fourier extensions in Section~\ref{sec:sobolev_regularised_extensions}. 

For $f(x)=x$, the rescaled fitted function from Eq.~\eqref{eq:g_definition} is $g(x)=\mu +\eta\delta x/\pi$. There is no parameter-dependent shrinking length scale, so we take $r_{\rm id}=1$. On $\Omega_\eta$,
\begin{equation}
    \|g\|_{L^2(\Omega_\eta)}^2=\frac{2\pi}{\eta}\left(\mu^2+\frac{\delta^2}{3}\right),\quad \|g'\|_{L^2(\Omega_\eta)}^2=\frac{2\eta}{\pi}\delta^2.
    \label{eq:identity_interval_norms}
\end{equation}
Since $\|\mathcal H\|=|\mu|+\delta$, Eq.~\eqref{eq:identity_interval_norms} gives $\|g\|_{H^1(\Omega_\eta)}=O(\|\mathcal H\|)$ for fixed $\eta>1$. Extending $g$ periodically using the fixed Gevrey cutoff using the proof of Theorem~\ref{thm:near_exponential_convergence} gives an extension $h_{\rm id}$ satisfying
\begin{equation}
    \|h_{\rm id}\|_{H^1(\mathbb T)}=O(\|\mathcal H\|),\quad
    \|h_{\rm id}^{(\ell)}\|_{L^1(\mathbb T)}
    \leq C_0^{\ell+1}(\ell!)^\sigma\|\mathcal H\|,
\end{equation}
for $\ell\geq1$, where $C_0$ depends only on $\sigma$, $\eta$, and $\eta'$. Thus $h_{\rm id}$ is scale-adapted at $r_{\rm id}=1$, and Lemma~\ref{lem:scale_adapted_extension} gives $C_f^{\rm id}=O(1)$ and hence $\alpha_m^{\rm reg}=O(\|\mathcal H\|)$.

\subsection{Matrix exponentiation}
\label{sec:matrix_exponentiation}

For $f(x)=e^x$, the normalized scalar function is $g(x) = \exp(\mu+\eta\delta x/\pi)$. Since $e^x$ is convex and endpoint-maximizing, the reflected endpoint construction of Section~\ref{sec:optimal_subnormalization_on_extended_domains} saturates the subnormalization lower bound, giving $\alpha_\infty^{\rm refl}=e^{\lambda_{\rm max}}=\|e^{\mathcal H}\|$ with algebraic convergence $\epsilon_m^{\rm refl}=O(m^{-1})$.

The arcsine-Taylor construction of Section~\ref{sec:arcsine_taylor_identity_function_decomposition} is also straightforward because $e^x$ is entire. Large values of $\eta$ may be chosen freely to improve the exponential convergence factor $\rho_\eta^{\rm asin} = \sin(\pi/\eta)$ in Eq.~\eqref{eq:arcsine_taylor_convergence} at a known cost to the subnormalization bound in Eq.~\eqref{eq:arcsine_taylor_subnormalization}. The convergence is $\epsilon_m^{\rm asin}=O(m^{-3/2}\sin^m[\pi/\eta])$, while the arcsine-Taylor asymptotic subnormalization and its ratio to the optimal value are
\begin{equation}
    \alpha_\infty^{\rm asin} = \exp\!\left(\mu+\frac{\eta\delta}{2}\right), \quad \frac{\alpha_\infty^{\rm asin}}{\alpha_\infty^{\rm refl}} = \exp\!\left( \left[\frac{\eta}{2}-1\right]\delta \right).
\end{equation}
The subnormalization is larger than the optimum $\alpha_\infty^{\rm refl} = f(\mu+\delta)$ achieved by the reflected extensions because the arcsine-Taylor endpoint is displaced from $\lambda_{\rm max} = \mu+\delta$ to $\mu+\eta\delta/2$. The subnormalization ratio shows that the arcsine-Taylor construction for $f(x)=e^x$ has an exponential dependence on the spectral half-width $\delta$. Therefore, when $\mathcal H$ has a broad spectrum and subnormalization is the dominant cost, the reflected-extension approach is preferable despite only converging algebraically. For narrow spectra, this overhead is mild, and the exponential convergence of the arcsine-Taylor construction will often be preferred. 

The Sobolev-regularized construction of Section~\ref{sec:sobolev_regularised_extensions} instead converges as $\epsilon_m^{\rm reg}=O(\exp[-C_1m^{1/\sigma}]+\epsilon_{\rm t})$ for every fixed $\sigma>1$. Since $e^x$ is entire, we recommend the fixed choice $\eta=2$, since this attains optimal subnormalization for reflected extensions. We have $g'(x)=(\eta\delta/\pi)g(x)$ and
\begin{equation}
    \begin{split}
        \|g\|_{L^2(\Omega_\eta)}^2  &=  \frac{\pi}{2\eta\delta} e^{2\lambda_{\rm max}}(1-e^{-4\delta}), \\
        \|g'\|_{L^2(\Omega_\eta)}^2 &= \left(\frac{\eta\delta}{\pi}\right)^2\|g\|_{L^2(\Omega_\eta)}^2.
    \end{split}
\end{equation}
The unweighted $H^1$-norm from Definition~\ref{def:weighted_sobolev_norm} therefore has an $O([\eta\delta]^{1/2})$ overhead above the optimal value, which would be transferred to the subnormalization using this regularizer. This can be eliminated by using the weighted Sobolev norm in Definition~\ref{def:weighted_sobolev_norm} with weight
\begin{equation}
    r_{\exp}=\left(1+\frac{\eta\delta}{\pi}\right)^{-1}.
\end{equation}
The choice of $r_{\exp}$ matches the characteristic variation scale of $g$, since $(\eta\delta/\pi)r_{\exp}\leq1$. Extending $g$ periodically using Gevrey transition layers of width $\Theta(r_{\exp})$ gives an extension $h_{\exp}$ satisfying Eq.~\eqref{eq:scale_adapted_condition}, where $C_0$ is independent of the spectral width. Thus $h_{\exp}$ is scale-adapted at $r_{\exp}$, and Lemma~\ref{lem:scale_adapted_extension} gives
\begin{equation}
    C_f^{\exp}=O(1),\qquad \alpha_m^{\rm reg}=O\!\left(\|e^{\mathcal H}\|\right).
    \label{eq:regularized_exponential_subnormalization}
\end{equation}

The Hermitian dilation does not allow for the exponentiation of general non-Hermitian matrices $A$, as exponentiation relies on eigenvalue, rather than singular-value, transforms. A route to extending the approach to non-Hermitian matrix exponentiation is operator splitting~\cite{Blanes2024}. For example, $A = \mathcal H_1 + i\mathcal H_2$ may be separated into its Hermitian $\mathcal H_1$ and anti-Hermitian $i\mathcal H_2$ parts~\cite{Brearley2025}, then evolved individually over $t/\Delta t \in \mathbb N$ steps of size $\Delta t$ as
\begin{equation}
    e^{At} =e^{(\mathcal H_1+i\mathcal H_2)t} = \left(e^{\mathcal H_1\Delta t}e^{i\mathcal H_2\Delta t}\right)^{t/\Delta t} + O(t\Delta t),
    \label{eq:operator_splitting}
\end{equation}
assuming $\|A\|\leq 1$. The $e^{\mathcal H_1\Delta t}$ factor is an imaginary-time evolution and can be block encoded using the present Fourier-extension method, while $e^{i\mathcal H_2\Delta t}$ is a Hamiltonian simulation. Splitting the time evolution into shorter steps has the effect of multiplying the subnormalization across each step, so the optimal-subnormalization construction in Section~\ref{sec:optimal_subnormalization_on_extended_domains} avoids an accumulating subnormalization penalty~\cite{Fang2023}.

\subsection{Matrix inversion}
\label{sec:matrix_inversion}

For inversion, the normalized fitted scalar function is
\begin{equation}
    g(x) = \left(\mu+\frac{x}{\tau}\right)^{-1} = \frac{\pi}{\pi\mu + \eta\delta x},
    \label{eq:g_inversion}
\end{equation}
so the pole of $f$ at $x=0$ is mapped to
\begin{equation}
    x_{\mathrm{pole}}=-\tau\mu = -\frac{\pi}{\eta}\frac{\mu}{\delta}.
    \label{eq:inverse_pole}
\end{equation}

Since $x^{-1}$ is convex and endpoint-maximizing on a positive interval, the reflected endpoint construction of Section~\ref{sec:optimal_subnormalization_on_extended_domains} is norm-saturating. Therefore,
\begin{equation}
    \alpha_\infty^{\rm refl} = \|\mathcal H^{-1}\| = \lambda_{\rm min}^{-1},
\end{equation}
while the convergence remains algebraic. Under the normalization $\operatorname{spec}(\mathcal H)\subseteq[\kappa^{-1},1]$, the endpoint derivative results in the slow convergence of $\epsilon_m^{\rm refl}=O(\kappa^2m^{-1})$, which gives an overall $O(\kappa^3)$ condition-number dependence for this method. However, since the condition-number dependence is concentrated in the algebraic convergence with the optimal subnormalization dependence, the approach may be competitive when larger values of $\epsilon$ are tolerable. The full resource requirements are discussed in Section~\ref{sec:resource_requirements}.

The arcsine-Taylor construction instead evaluates the asymptotic subnormalization at the enlarged maximizing endpoint. Assuming a positive spectrum, this gives 
\begin{equation}
    \alpha_\infty^{\rm asin} = \left(\mu-\frac{\eta\delta}{2}\right)^{-1}, \qquad 2<\eta<\frac{2\mu}{\delta}.
    \label{eq:inverse_arcsine_alpha}
\end{equation}
The exponential convergence factor is $\rho_\eta^{\rm asin} = \sin(\pi/\eta)$, but the admissible values of $\eta$ are restricted by the pole at zero. The ratio relative to the reflected, asymptotically optimal value is
\begin{equation}
    \frac{\alpha_\infty^{\rm asin}}{\alpha_\infty^{\rm refl}} = \frac{\lambda_{\rm min}}{\mu-\eta\delta/2} = \frac{4}{2(\kappa+1)-\eta(\kappa-1)}.
    \label{eq:inverse_arcsine_ratio}
\end{equation}
This gives optimal subnormalization at $\eta=2$, but diverges as $\eta$ approaches the analyticity bound $2\mu/\delta = 2(\kappa+1)/(\kappa-1)$. Thus inversion presents a particularly sharp trade-off, where increasing $\eta$ improves the exponential convergence factor $\rho_\eta^{\rm asin}=\sin(\pi/\eta)$, but also drives the enlarged interval towards the pole at zero and can dramatically increase the asymptotic subnormalization. For large condition numbers $\kappa$, the admissible range of $\eta$ collapses towards $2$ where the convergence becomes algebraic, so the advantage of the arcsine-Taylor construction becomes correspondingly limited. 

For Sobolev-regularized Fourier extensions, Eq.~\eqref{eq:inverse_pole} gives a useful way of choosing the extension factor. Placing the pole at the periodic boundary gives 
\begin{equation}
    \eta_{\mathrm{def}} = \frac{\kappa+1}{\kappa-1}
    \label{eq:eta_one_interval}
\end{equation}
for a positive-definite $\mathcal H$. If $\mathcal H$ is indefinite and nonsingular, then no single rescaled interval can both avoid the pole at zero and contain the whole spectral interval. In this case, we fit on a disconnected spectral set. Let the singular values $\operatorname{sv}(\mathcal H)\subseteq[\Sigma_{\rm min},\Sigma_{\rm max}]$, with $\kappa=\Sigma_{\rm max}/\Sigma_{\rm min}$. Since the spectrum is naturally centered, use $\mathcal G = \tau\mathcal H$ for $\tau = \pi/(\eta \Sigma_{\rm max})$ so that the positive and negative parts of the spectrum are mapped into
\begin{equation}
    \Omega_{\eta,\kappa}^{\rm indef} =
    [-x_{\rm max},-x_{\rm min}]\cup[x_{\rm min},x_{\rm max}],
    \label{eq:two_interval_inverse_domain}
\end{equation}
where $x_{\rm max}=\pi/\eta$ and $x_{\rm min}=\pi/(\kappa\eta)$. On this set, $g(x)=\tau/x$ is odd, so a sine-series representation is natural. Matching the distance from the inner endpoint to the pole at zero with the distance from the outer endpoint to the periodic boundary gives
\begin{equation}
    \eta_\mathrm{indef} = 1+\frac{1}{\kappa}.
    \label{eq:eta_two_interval}
\end{equation}

The natural length scale is set by the gap to the nearest pole. For the definite and indefinite fits, respectively, we take
\begin{equation}
    r_{\rm def}=\min\left\{1,\frac{2\pi}{\kappa+1}\right\},\;\; r_{\rm indef}=\min\left\{1,\frac{\pi}{\kappa+1}\right\}.
\end{equation}
Under the normalization $\operatorname{sv}(\mathcal H)\subseteq[\kappa^{-1},1]$, we have $r^{-1}=\Theta(\kappa)$ and $\|\mathcal H^{-1}\|=\Theta(\kappa)$. This gives 
\begin{equation*}
    \|g\|_{L^2(\Omega)}=O\!\left(r^{\frac{1}{2}}\|\mathcal H^{-1}\|\right),\; r\|g'\|_{L^2(\Omega)}=O\!\left(r^{\frac{1}{2}}\|\mathcal H^{-1}\|\right)\!,
\end{equation*}
and $\|g^{(\ell)}\|_{L^1(\Omega)}=O(\ell!r^{1-\ell}\|\mathcal H^{-1}\|)$ for $\ell\geq1$. Gevrey transition layers of width $\Theta(r)$ preserve these scalings, giving a periodic extension $h_{\rm inv}$ satisfying Eq.~\eqref{eq:scale_adapted_condition} where $C_0$ is independent of $\kappa$. Thus $h_{\rm inv}$ is scale-adapted at scale $r$, and Lemma~\ref{lem:scale_adapted_extension} gives
\begin{equation}
    C_f^{\rm inv}=O(1),\qquad \alpha_m^{\rm reg}=O\!\left(\|\mathcal H^{-1}\|\right).
    \label{eq:regularized_inverse_subnormalization}
\end{equation}

Replacing $\mathcal H$ by the Hermitian dilation $\mathcal H(A)$ gives
\begin{equation}
    \operatorname{spec}(\mathcal H(A))=\pm\operatorname{sv}(A),
\end{equation}
so the dilation is automatically indefinite but nonsingular whenever $A$ is nonsingular. Its inverse is
\begin{equation}
    \mathcal H(A)^{-1} = 
    \begin{pmatrix}
        0 & A^{-1}\\
        \left(A^\dagger\right)^{-1} & 0
    \end{pmatrix},
\label{eq:hermitian_dilation_inversion}
\end{equation}
so a block encoding of $\mathcal H(A)^{-1}$ yields a block encoding of $A^{-1}$.

\subsection{Matrix square root}
\label{sec:matrix_square_root}

For $f(x)=\sqrt{x}$, we assume throughout this example that $\mathcal H=\mathcal H^\dagger$ is positive definite with spectrum $\operatorname{spec}(\mathcal H) \subseteq[\lambda_{\rm min}, \lambda_{\rm max}]$ with $\lambda_{\rm min}>0$. While concave functions like $\sqrt{x}$ do not satisfy the condition of Lemma~\ref{lem:convex_endpoint_reflection_saturation} for optimal subnormalization, the subnormalization remains bounded independently of the error for every fixed positive spectral interval. The convergence is algebraic as $\epsilon_m^{\rm refl}=O(m^{-1})$, with constants depending on the separation from the branch point at the origin. 

Since asymptotic optimality has already been compromised, it is worthwhile to use the arcsine-Taylor construction to improve the convergence to exponential. Theorem~\ref{thm:taylor_sine_lcu} guarantees bounded subnormalization together with $\epsilon_m^{\rm asin}=O(m^{-3/2}\sin^m[\pi/\eta])$ for every fixed admissible $\eta$, which is $2<\eta<2\mu/\delta$ to prevent the branch point from approaching the interval. When $\eta$ is increased, the approximation is performed over the enlarged effective interval $\Omega_{\eta}^{\rm asin} = [\mu-\eta\delta/2, \mu+\eta\delta/2]$, so improving the convergence factor $\rho_{\eta}^{\rm asin} = \sin(\pi/\eta)$ comes at the cost of a larger subnormalization scale. For concave, monotonically increasing functions such as $\sqrt{x}$, this trade-off is often favorable, since the increase in the subnormalization scale is tied to the value of the function at the enlarged maximizing endpoint $\mu+\eta\delta/2$.

\begin{figure*}
    \centering
    \includegraphics[width=0.99\textwidth]{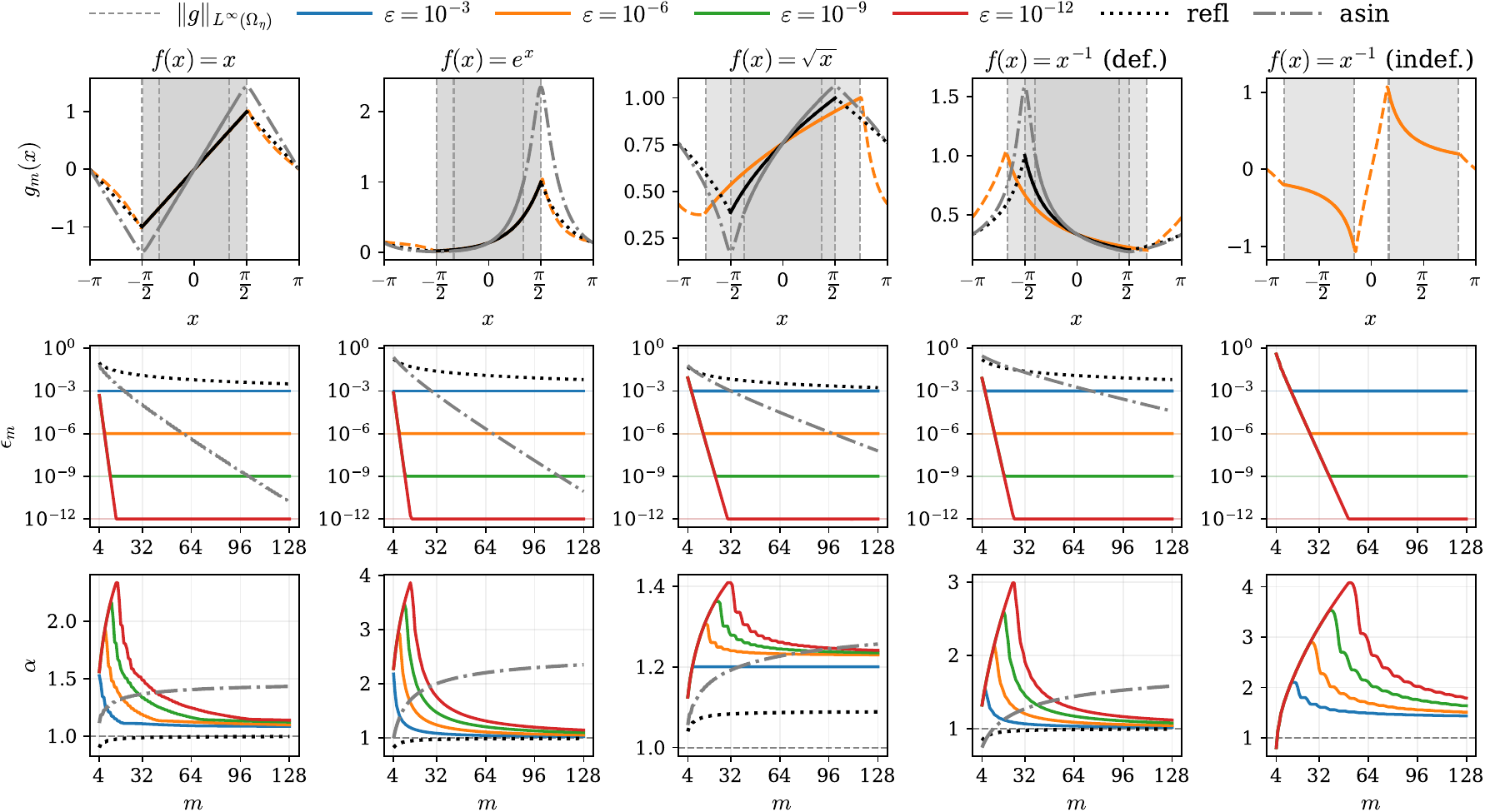}
    \caption{Approximations of the identity, exponential, square-root, and inverse functions for the Sobolev-regularized (``reg''), arcsine-Taylor (``asin''), and reflected-extension (``refl'') Fourier extensions. Top row: Fourier-extension approximation $g_m(x)$ for $m=128$, where the fitted spectral intervals are shown by the shaded regions and solid lines. Middle row: approximation error $\epsilon_m$ in Eq.~\eqref{eq:spectral_error} against $m$. Bottom row: LCU subnormalization in Eq.~\eqref{eq:alpha_definition} against $m$ with lower bound set by $\|f(\mathcal H)\|=1$.}
    \label{fig:fourier_extensions}
\end{figure*}

For the Sobolev-regularized extensions, the robust choice $\eta=(\kappa+1)/(\kappa-1)$ from Eq.~\eqref{eq:eta_one_interval} places the branch point at the periodic boundary. Unlike inversion, the closure width shrinks with $\kappa$ without compensating growth in $\|\sqrt{\mathcal H}\|$. The complement of the fitted interval has width $l_\kappa=4\pi/(\kappa+1)=\Theta(\kappa^{-1})$, while the periodic extension must bridge the endpoint difference $\Delta_\kappa=\sqrt{\lambda_{\rm max}}-\sqrt{\lambda_{\rm min}}=\Theta(\sqrt{\lambda_{\rm max}})$. Since $h=g$ on $\Omega_\eta$, the $L^2$ contribution to $\|h\|_{H_r^w(\mathbb T)}^2$ is $\Omega(\lambda_{\rm max})$. For fixed $1/2<w\leq 1$, bridging a difference $\Delta_\kappa$ across an interval of width $l_\kappa$ contributes at least $\Omega(r^{2w}\Delta_\kappa^2l_\kappa^{1-2w})$ to the weighted Sobolev norm. A Gevrey interpolation rescaled to the complementary interval attains the corresponding upper bound, so an appropriate periodic extension satisfies
\begin{equation}
    \|h\|_{H_r^w(\mathbb T)}^2=\Theta\!\left(\lambda_{\rm max}\left[1+r^{2w}\kappa^{2w-1}\right]\right).
\end{equation}
Lemma~\ref{thm:l1h1} and Eq.~\eqref{eq:Cf_extension_constant} then give
\begin{equation}
    C_f^{\rm sqrt}=\Theta\!\left(r^{-1/2}\sqrt{1+r^{2w}\kappa^{2w-1}}\right).
\end{equation}
Balancing these two contributions gives $r_{\rm sqrt}=\Theta(\kappa^{-(2w-1)/(2w)})$, and hence
\begin{equation}
    C_f^{\rm sqrt}=\Theta\!\left(\kappa^{\frac{2w-1}{4w}}\right),\quad \alpha_m^{\rm reg}=O\!\left(\kappa^{\frac{2w-1}{4w}}\sqrt{\lambda_{\rm max}}\right).
    \label{eq:regularized_sqrt_fractional_subnormalization}
\end{equation}
Thus, unlike inversion, the shrinking periodic closure is not compensated by growth in the operator norm, and $C_f^{\rm sqrt}$ cannot remain $O(1)$ for any fixed $w>1/2$. The exponent approaches zero as $w\downarrow1/2$, but the coefficient-bound constant $C_w$ diverges in this limit. For the $H_r^1$ regularizer, this reduces to $r_{\rm sqrt}=\kappa^{-1/2}$, $C_f^{\rm sqrt}=\Theta(\kappa^{1/4})$, and $\alpha_m^{\rm reg}=O(\kappa^{1/4}\sqrt{\lambda_{\rm max}})$.

\subsection{Numerical comparison}
\label{sec:numerical_comparison}

We now compare the implementations across all three coefficient selection strategies: norm-saturating reflected extensions from Section~\ref{sec:optimal_subnormalization_on_extended_domains}, exponentially convergent arcsine-Taylor extensions from Section~\ref{sec:exponential_convergence_with_bounded_subnormalization}, and Sobolev-regularized extensions from Section~\ref{sec:sobolev_regularised_extensions}. Each strategy requires the same standard LCU block encoding circuits~\cite{Childs2012}, differing only by the coefficients. The primary quantities of interest are the spectral error in Eq.~\eqref{eq:spectral_error} and LCU subnormalization in Eq.~\eqref{eq:alpha_definition}.

Figure~\ref{fig:fourier_extensions} shows the three coefficient-selection strategies applied to the identity, exponential, square-root, definite inverse, and indefinite inverse examples. In all cases, the fitted spectral intervals are chosen so that the operator-norm lower bound on the subnormalization is $\|f(\mathcal H)\|=1$. For $f(x)=x$, $e^x$, and $\sqrt{x}$, we use spectral intervals $\operatorname{spec}(\mathcal H)\subset[\lambda_{\rm min}, \lambda_{\rm max}]$ that are chosen tightly within $[-1,1]$, $[-4,0]$, and $[0.15,1]$, respectively. For matrix inversion, we consider both definite $[1,5]$ and indefinite $[-5,5]$ spectra. The indefinite inverse is the only example requiring a disconnected fitted region, here given by $[-5,-1]\cup[1,5]$.

The reflected extensions use $\eta=2$ throughout. This gives optimal subnormalization for the identity, exponential, and inverse examples that support the convexity requirement in Lemma~\ref{lem:convex_endpoint_reflection_saturation}. The square root is the exception, since it is concave and does not satisfy the endpoint-saturation condition. The middle row of Fig.~\ref{fig:fourier_extensions} shows the corresponding limitation of reflected extensions, that the convergence is only first order, $\epsilon_m=O(m^{-1})$, and reaches errors only of the order of approximately $10^{-2}$ by $m=128$ for these examples.

The arcsine-Taylor curves show the expected improvement to exponential convergence, at the cost of a larger limiting subnormalization. For entire functions, we choose $\eta=3$, giving $\rho_\eta^{\rm asin}=\sin(\pi/3)\approx0.866$. This is visible in the identity and exponential error curves, while the bottom row shows the associated subnormalization overhead. For non-entire functions, $\eta$ is restricted by the nearest pole or branch point. For the square root and definite inverse examples, we require $\eta<2\mu/\delta$. In Fig.~\ref{fig:fourier_extensions}, this gives the upper bounds $46/17\approx2.706$ for the square root and $3$ for the definite inverse, so we use $\eta=2.7$ and $\eta=2.5$, respectively. 

\begin{figure*}[t]
    \centering
    \includegraphics[width=0.99\textwidth]{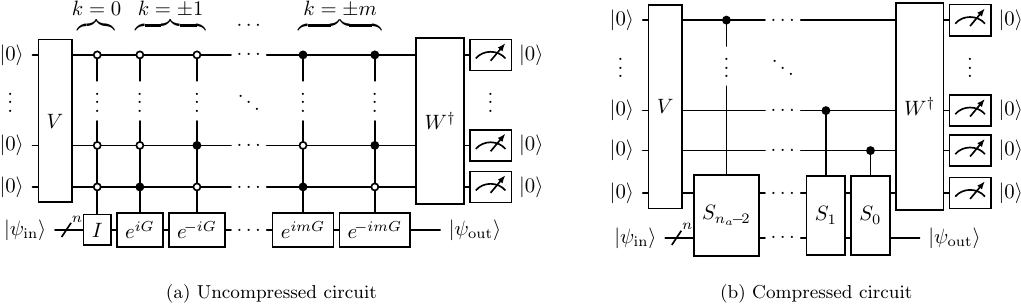}
    \caption{LCU circuits for block encoding matrix functions from Hamiltonian-simulation unitaries $U_k=e^{ik\mathcal G}$. The ancilla unitaries $V$ and $W$ encode the coefficient magnitudes and phases, as defined in Eqs.~\eqref{eq:V_definition} and~\eqref{eq:W_definition}. (a) Uncompressed circuit in the standard LCU formulation, where each Fourier mode is a separate controlled branch. (b) Compressed binary-power implementation of the same LCU block encoding, where $S_j$ is defined in Eq.~\eqref{eq:Sj_definition}.}
    \label{fig:circuit}
\end{figure*}

The Sobolev-regularized curves use $w=1$ and converge rapidly to the prescribed target tolerances $\epsilon_{\rm t}$. Numerically, the observed convergence is consistent with $\rho_\eta^{\rm reg} = \tan^2(\pi/[4\eta])$, consistent with unregularized $L^2$ convergence~\cite{Adcock2014resolution}. Using $\eta=2$ gives $\rho_2^{\rm reg}\approx 0.172$ for all entire $f$. For square root and inversion, $\eta$ is chosen using the condition-number dependent model of Eq.~\eqref{eq:eta_one_interval}, so that the branch point or pole is placed at the periodic boundary rather than close to the fitted endpoint. The bottom row shows that the regularized fits initially pay a subnormalization penalty while resolving the target tolerance, before relaxing towards a value close to the operator-norm lower bound. Because the regularized problem minimizes a smooth Sobolev $\ell^2$-type surrogate rather than the plotted coefficient $\ell^1$-norm $\alpha$, independently computed solutions at successive $m$ do not necessarily have monotonically decreasing subnormalization. We therefore retain the lowest-$\alpha$ feasible solution obtained up to each $m$, giving the non-smooth appearance in some cases.

\section{Compressed LCU circuit}
\label{sec:quantum_circuit}

Figure~\ref{fig:circuit} presents the quantum circuits for block encoding analytic functions of Hermitian matrices $f(\mathcal H)$. Figure~\hyperref[fig:circuit]{\ref*{fig:circuit}a} shows the uncompressed quantum circuit in the standard LCU formulation where each unitary is implemented explicitly. Figure~\hyperref[fig:circuit]{\ref*{fig:circuit}b} shows a compressed equivalent, which is possible since all unitaries are the same Hamiltonian simulation $e^{ik\mathcal G}$ evolved across integer evolution times $k\in\mathbb Z$, which may be written as
\begin{equation}
    |k| = \sum_{j=0}^{n_{\rm a}-2} r_j 2^j,\qquad
    r_j\in\{0,1\}
    \label{eq:signed_binary_magnitude}
\end{equation}
for qubit $r_j$. Substituting this expression into the unitary $e^{ik\mathcal G}$ 
gives

\begin{equation}
    e^{ik\mathcal G}=\exp\left(\pm i\mathcal G\sum_{j=0}^{n_{\rm a}-2}r_j2^j\right)=\prod_{j=0}^{n_{\rm a}-2}\exp\left(\pm i2^jr_j\mathcal G\right).
    \label{eq:circuit_compression}
\end{equation}
This simplification reveals that the evolution can be implemented by a product of the simple exponentials conditional on qubit $r_j$. The signs are implemented by defining the unitary
\begin{equation}
    S_j=\ket{0}\!\bra{0}\otimes e^{i2^j\mathcal G}+\ket{1}\!\bra{1}\otimes e^{-i2^j\mathcal G},
    \label{eq:Sj_definition}
\end{equation}
where the least-significant ancilla qubit in Fig.~\hyperref[fig:circuit]{\ref*{fig:circuit}b} acts as the sign qubit. This is applied with successive controls as determined by Eq.~\eqref{eq:circuit_compression}, as shown in Fig.~\ref{fig:circuit}. All ancilla qubits are naturally zero at the index corresponding to the zero mode, so the Hamiltonian simulation is naturally not applied.

The LCU has $2m+1$ coefficients, so choosing $m$ as one less than a power of $2$ occupies the $2^{n_{\rm a}}$-dimensional ancilla register most effectively. The first column of $V$ is
\begin{equation}
    V_{j,0} = \sqrt{\frac{|\beta_j|}{\alpha}}.
    \label{eq:V_definition}
\end{equation}
The phase information is then encoded into the first column of a different unitary, $W$, as
\begin{equation}
    W_{j,0} = \frac{\beta_j^*}{|\beta_j|}\sqrt{\frac{|\beta_j|}{\alpha}}, \qquad \beta_j \neq 0,
    \label{eq:W_definition}
\end{equation}
with $W_{j,0}=0$ when $\beta_j=0$. This results in $V_{j,0}W_{j,0}^* = \beta_j/\alpha$. Unused rows in the first column are set to zero, and the remaining columns form an orthonormal basis so that $V$ and $W$ are unitary. The application of $V$ is a state preparation by its first column. The $W^\dagger$ operation combines the remaining magnitude and phase information to finalize the map
\begin{align}
    \ket{0}^{\otimes n_{\rm a}}\ket{\psi_{\rm in}}\mapsto\ket{0}^{\otimes n_{\rm a}}\frac{1}{\alpha}\sum_{j=0}^{2m}\beta_jU_j\ket{\psi_{\rm in}}+\dots,
    \label{eq:lcu_map}
\end{align}
as shown by \citeauthor{Childs2012}~\cite{Childs2012}. The indexing from $k=\{-m, \dots, m\}$ to $j = \{0, \dots, 2m\}$ uses the ordering $\{0, 1, -1, \dots, m, -m\}$, as shown in Fig.~\ref{fig:circuit}. Postselecting the ancilla qubits on $\ket{0}^{\otimes n_{\rm a}}$ yields the state $(A/\alpha)\ket{\psi}$, up to block-encoding error $\epsilon/\alpha$ in the postselected branch.

\section{Resource requirements}
\label{sec:resource_requirements}

\begin{table*}[t] 
    \setlength\tabcolsep{6pt}
    \def\arraystretch{2.2}
    \centering
    \caption{Resource scaling of the Fourier-extension block encodings with the coefficient selection strategies presented in Sections~\ref{sec:optimal_subnormalization_on_extended_domains}, \ref{sec:exponential_convergence_with_bounded_subnormalization}, and \ref{sec:sobolev_regularised_extensions}. Case-independent expressions in terms of $m$ are given in Eqs.~\eqref{eq:N_HS} and \eqref{eq:C_HS}, with the corresponding $m$-scaling substituted for the case dependence. The subnormalization for reflected extension and arcsine-Taylor assumes a convex endpoint-maximizing $f$. The Sobolev-regularized extensions set the internal target tolerance $\epsilon_{\rm t} = \epsilon$, and for entire $f$, assume $\eta=2$. Section~\ref{sec:application_to_various_functions} shows that $C_f=O(1)$ for the identity, exponential, and inverse functions, and $O(\kappa^{1/4})$ for the square-root function.}
    \label{tab:resource_scalings}
    \begin{tabular}{cccc}
    \toprule \\ [-10.5mm]
    & Reflected extension & Arcsine-Taylor & Sobolev-regularized \\[-1mm]
    \midrule
    $\epsilon_m$ & $O(m^{-1})$ & $O\!\left(m^{-3/2}\sin^m\!\left[\dfrac{\pi}{\eta}\right]\right)$ & $O\!\left(\exp[-C_1m^{1/\sigma}]+\epsilon_{\rm t}\right)$ \\
    $\alpha_\infty$ & $\|f(\mathcal H)\|$ & $\|f(\mathcal H)\| + O\left([\eta-2]\delta |f'(\lambda_\eta^\star)|\right)$ & $O(C_f\|f(\mathcal H)\|)$ \\
    $N_{\rm HS}$  
    & $O\!\left(
        \dfrac{\|f(\mathcal H)\|}{\|f_m(\mathcal H)|\psi\rangle\|_2}
        \log[\epsilon^{-1}]
      \right)$
    & $O\!\left(
        \dfrac{\alpha_m}{\|f_m(\mathcal H)|\psi\rangle\|_2}
        \log\!\left[
        \dfrac{\log(1/\epsilon)}
        {\log\!\csc(\pi/\eta)}
        \right]
      \right)$
    & $O\!\left(\dfrac{C_f\|f(\mathcal H)\|}{\|f_m(\mathcal H)\ket{\psi}\|_2}\log\!\log[\epsilon^{-1}]\right)$ \\
    $C_{\rm HS}$ 
    & $O\!\left(
        \dfrac{\|f(\mathcal H)\|}{\|f_m(\mathcal H)|\psi\rangle\|_2}
        s\epsilon^{-1}\log N
      \right)$
    & $O\!\left(
        \dfrac{\alpha_m}{\|f_m(\mathcal H)|\psi\rangle\|_2}
        s\log[\epsilon^{-1}]\log N
      \right)$
    & $O\!\left(
        \dfrac{C_f\|f(\mathcal H)\|}{\|f_m(\mathcal H)|\psi\rangle\|_2}
        s\log^\sigma[\epsilon_{\mathrm{t}}^{-1}]
        \log N
      \right)$ \\
    $N_{\rm HS}^{\rm inv}$ & $O\!\left(\kappa\log[\kappa\epsilon^{-1}]\right)$ & $O\!\left(\kappa[\log\kappa+\log\!\log(\kappa\epsilon^{-1})]\right)$ & $O\!\left(\kappa[\log\kappa+\log\!\log(\kappa\epsilon^{-1})]\right)$ \\
    $C_{\rm HS}^{\rm inv}$ & $O\!\left(s\kappa^3\epsilon^{-1}\log N\right)$ & $O\!\left(s\kappa^3\log[\kappa\epsilon^{-1}]\log N\right)$ & $O\!\left(s\kappa^2\log^\sigma[\kappa\epsilon^{-1}]\log N\right)$ \\
    \bottomrule
    \end{tabular}
\end{table*}

The cost of implementing $f(\mathcal H)$ on the quantum state can be quantified in terms of the gate complexity of (i) repeated coefficient loading by $V$ and $W^\dagger$ and (ii) the gate complexity of the controlled Hamiltonian simulations. If $C_{\rm VW}$ and $C_{\rm HS}$ are the respective two-qubit gate requirements, the gate complexity of preparing a state proportional to $f_m(\mathcal H)\ket{\psi}$ is $C_{\rm VW}+C_{\rm HS}$. As we see below, $C_{\rm VW}$ is absorbed into $C_{\rm HS}$ in the final expressions summarized in Table~\ref{tab:resource_scalings}. We also quantify the scaling in terms of the number of uses of the Hamiltonian simulation algorithm $N_{\rm HS}$ and the required simulation tolerance $\epsilon_{\rm HS}$ to achieve a final state accurate to $\epsilon_\psi$.

For the two-qubit gate counts, we use the standard sparse-access model for an $s$-sparse $N\times N$ Hamiltonian. An optimal sparse Hamiltonian simulation algorithm~\cite{Low2017, Low2019} simulates an $s$-sparse Hamiltonian $\mathcal G$ for time $t$ with query complexity $O(s t\|\mathcal G\|+\log[\epsilon_{\rm HS}^{-1}]/\log\!\log[\epsilon_{\rm HS}^{-1}])$, conservatively writing $\|\mathcal G\|_{\max}\leq\|\mathcal G\|$. Assuming the sparse-access oracles and arithmetic are implemented efficiently, each query contributes $O(\log N)$ two-qubit gates, up to lower-order polylogarithmic factors. This analysis is comparable to the sparse Hamiltonian-simulation-based QLSA analysis of \citeauthor{Childs2017}~\cite{Childs2017}. In the following subsection, we keep the leading dependence on $s$, $\log N$, and the total simulated time, while suppressing the lower-order precision polylogarithms.

\subsection{In terms of Fourier truncation order \texorpdfstring{\boldmath$m$}{m}}

The $V$ and $W^\dagger$ unitaries act on $n_{\rm a} = \lceil\log_2(2m+1)\rceil = O(\log m)$ qubits, and since we only require unitaries with the first columns of $V$ and $W$, arbitrary $n_{\rm a}$-qubit state preparation requires $O(2^{n_{\rm a}}) = O(m)$ two-qubit gates~\cite{Shende2006}. The number of calls to each of $V$ and $W^\dagger$ is the number of calls to the block encoding. A single execution of the circuits in Fig.~\ref{fig:circuit} followed by measuring the ancilla qubits to be $\ket{0}^{\otimes n_{\rm a}}$ succeeds with probability
\begin{equation}
    p_m(\psi) = \frac{1}{\alpha_m^2}\|f_m(\mathcal H)\ket{\psi}\|_2^2.
    \label{eq:probability}
\end{equation}
Using amplitude amplification, attaining $p_m = O(1)$ requires $O(p_m(\psi)^{-1/2}) = O(\alpha_m / \|f_m(\mathcal H)\ket{\psi}\|_2 )$ calls to the block encoding~\cite{Brassard2002}. The gate complexity of coefficient loading by $V$ and $W^\dagger$ is therefore 
\begin{equation}
    C_{\rm VW} = O\!\left( \frac{\alpha_m}{\|f_m(\mathcal H)\ket{\psi}\|_2} m \right).
\end{equation}

The compressed circuit in Fig.~\hyperref[fig:circuit]{\ref*{fig:circuit}b} requires $O(\log m)$ Hamiltonian simulations per block encoding, which gives
\begin{equation}
    N_{\rm HS} = O\!\left( \frac{\alpha_m}{\|f_m(\mathcal H)\ket{\psi}\|_2} \log m \right)
    \label{eq:N_HS}
\end{equation}
calls to the Hamiltonian simulation algorithm when combined with amplitude amplification~\cite{Brassard2002}. Each unitary $S_j$ defined in Eq.~\eqref{eq:Sj_definition} implements the signed simulation $e^{\pm i2^j\mathcal G}$. If $\mathcal H$ is $s$-sparse, $\mathcal G=\tau(\mathcal H-\mu I)$ is $O(s)$-sparse, and $\operatorname{spec}(\mathcal G)\subseteq[-\pi/\eta,\pi/\eta]$. The leading sparse-query cost of implementing $e^{\pm i2^j\mathcal G}$ is therefore $O(s2^j\eta^{-1})$. Summing the binary powers gives a leading sparse-query cost $O(sm\eta^{-1})$, and hence a two-qubit gate cost $O(sm\eta^{-1}\log N)$ under the oracle-efficiency assumptions above. After amplitude amplification, the two-qubit gate complexity of the controlled Hamiltonian simulation's contribution is
\begin{equation}
    C_{\rm HS} = O\!\left(\frac{\alpha_m}{\|f_m(\mathcal H)\ket{\psi}\|_2}\frac{sm}{\eta}\log N \right).
    \label{eq:C_HS}
\end{equation}

To prepare a state proportional to $f_m(\mathcal H)\ket{\psi}$ satisfying the error condition in Eq.~\eqref{eq:epsilon_psi}, each Hamiltonian simulation must be implemented to accuracy
\begin{equation}
    \epsilon_{\rm HS} = O\!\left( \frac{\epsilon_\psi}{N_{\rm HS}} \right) = O\!\left(\!\epsilon_\psi\frac{\|f_m(\mathcal H)\ket{\psi}\!\|_2}{\alpha_m\!\log m}\! \right)
    \label{eq:epsilon_psi_HS}
\end{equation}
where $\epsilon_\psi$ is the error in the final amplified state. The precision dependence associated with $\epsilon_{\rm HS}$ is included in the lower-order suppressed polylogarithmic factors. The overall gate complexity combining coefficient loading and Hamiltonian simulation is
\begin{equation}
    C_{\rm VW} + C_{\rm HS} = O\!\left( \frac{\alpha_m m}{\|f_m(\mathcal H)\ket{\psi}\|_2} \left[ 1 + \frac{s\log N}{\eta} \right]\right),
\end{equation}
where the coefficient-loading term is also of lower order, since in most settings, $s\eta^{-1}\log N =\Omega(1)$. The overall two-qubit gate complexity is then simply $C_{\rm HS}$.

We now derive the $m$ dependence for the various coefficient-selection strategies, and give specific examples for matrix inversion where there is an explicit dependence on the condition number $\kappa = \Sigma_{\rm max}/\Sigma_{\rm min}$, which is the ratio of the largest and smallest singular values of the input matrix $\mathcal H$. For consistency with the existing quantum linear systems algorithm literature~\cite{Harrow2009,Childs2017}, we normalize the singular-value interval so that $\operatorname{sv}(\mathcal H) \subseteq[\kappa^{-1}, 1]$. With this normalization, $\|\mathcal H^{-1}\|=O(\kappa)$ and $\|\mathcal H^{-1}\ket{\psi}\|_2=\Omega(1)$. Thus any construction with $\alpha_m=O(\|\mathcal H^{-1}\|)=O(\kappa)$ has worst-case amplitude-amplification cost $O(\kappa)$. The remaining condition-number dependence enters through the number of modes required to approximate $x^{-1}$ near the spectral endpoint $\Sigma_{\rm min}$. 

\subsection{Resources for reflected extensions}

For the reflected extensions of Section~\ref{sec:optimal_subnormalization_on_extended_domains}, $\eta=2$ is fixed and the subnormalization is optimal, but the convergence is algebraic
\begin{equation}
    \epsilon_m^{\rm refl} =O(m^{-1}), \qquad
    \alpha_m^{\rm refl}=\|f(\mathcal H)\|+O(m^{-1}),
\end{equation}
assuming the fitted interval is spectrally tight, so $m=O(\epsilon^{-1})$. This approach therefore requires 
\begin{equation}
    N_{\rm HS}^{\rm refl} = O\!\left( \frac{\|f(\mathcal H)\|}{\|f_m(\mathcal H)\ket{\psi}\|_2} \log [\epsilon^{-1}] \right)
\end{equation}
calls to the Hamiltonian simulation algorithm, and
\begin{equation}
    C_{\rm HS}^{\rm refl} = O\!\left( \frac{\|f(\mathcal H)\|}{\|f_m(\mathcal H)\ket{\psi}\!\|_2}s\epsilon^{-1}\log N\right),
    \label{eq:complexity_algebraic}
\end{equation}
two-qubit gates. This approach minimizes the amplitude-amplification overhead, but is only algebraic in the precision.

For matrix inversion by $f(x)=x^{-1}$, the endpoint derivative scales as $O(\kappa^2)$, giving
\begin{equation}
    \epsilon_m^{\rm alg,inv} = O\!\left(\kappa^2m^{-1}\right), \qquad
    m = O\!\left(\kappa^2\epsilon^{-1}\right).
\end{equation}
Since $\alpha_m^{\rm alg,inv} = O(\kappa)$, the reflected extensions method requires
\begin{equation}
    N_{\rm HS}^{\rm alg,inv} = O\!\left( \kappa \log [\kappa\epsilon^{-1}] \right)
\end{equation}
calls to the Hamiltonian simulation and
\begin{equation}
    C_{\rm HS}^{\rm alg,inv} = O\!\left(s\kappa^3\epsilon^{-1}\log N \right),
    \label{eq:reflected_extensions_inverse_complexity}
\end{equation}
two-qubit gates.

\subsection{Resources for arcsine-Taylor}

For the arcsine-Taylor construction of Section~\ref{sec:exponential_convergence_with_bounded_subnormalization}, the error satisfies
\begin{equation}
    \epsilon_m^{\rm asin} = O\!\left( m^{-3/2}\sin^m[\pi/\eta] \right), \qquad \eta>2.
\end{equation}
Solving for $m$ gives
\begin{equation}
    m = O\!\left( \frac{\log(1/\epsilon)} {\log\!\csc(\pi/\eta)} \right),
    \label{eq:m_arcsine_resource}
\end{equation}
for fixed $\eta$. Assuming a fixed $\eta>2$, the arcsine-Taylor method requires
\begin{equation}
    N_{\rm HS}^{\rm asin} = O\!\left( \frac{\alpha_\infty^{\rm asin}}{\|f_m(\mathcal H)\ket{\psi}\|_2} \log\!\log[\epsilon^{-1}] \right)
\end{equation}
calls to the Hamiltonian simulation, and
\begin{equation}
     C_{\rm HS}^{\rm asin} = O\!\left( \frac{\alpha_\infty^{\rm asin}}{\|f_m(\mathcal H)\ket{\psi}\|_2} \frac{s}{\eta}\log N\frac{\log[\epsilon^{-1}]} {\log\!\csc[\pi\eta^{-1}]} \right)
    \label{eq:complexity_arcsine}
\end{equation}
two-qubit gates. The expression may be further simplified by considering a fixed $\eta$, which is the form given in Table~\ref{tab:resource_scalings}.

For matrix inversion, the constraints on $\eta$ in Eq.~\eqref{eq:inverse_arcsine_alpha} retain the optimal $O(\kappa)$ amplitude amplification factor, and $\eta=2+\Theta(\kappa^{-1})$ which gives $\log\!\csc(\pi \eta^{-1})=\Theta(\kappa^{-2})$. The required number of modes is therefore
\begin{equation}
    m = O\!\left( \kappa^2\log\!\left[\frac{\kappa}{\epsilon}\right] \right).
\end{equation}
This requires
\begin{equation}
    N_{\rm HS}^{\rm asin} = O\!\left( 
    \kappa \log \left[\kappa^2\log\!\left(\frac{\kappa}{\epsilon}\right)\right] \right)
\end{equation}
calls to the Hamiltonian simulation, and
\begin{equation}
    C_{\rm HS}^{\rm asin,inv} = O\!\left( s\kappa^3 \log\!\left[\frac{\kappa}{\epsilon}\right]\log N \right).
    \label{eq:arcsine_inverse_complexity}
\end{equation}
two-qubit gates. 

For convex functions whose maximum is attained at an endpoint, Corollary~\ref{cor:endpoint_saturated_asymptotic_subnormalization} gives the asymptotic bound for convex endpoint-maximizing functions
\begin{equation}
    \alpha_\infty^{\rm asin} = \left| f\!\left(\mu+q\frac{\eta\delta}{2}\right) \right|, \qquad
    q\in\{-1,1\},
    \label{eq:arcsine_alpha_resource}
\end{equation}
where the sign $q$ selects the endpoint at which $|f|$ is largest. The subnormalization may be written as $\alpha_\infty = |f(\lambda_\eta^\star)|$ where $\lambda_\eta^\star = \mu+q\eta\delta/2$ is the endpoint of the enlarged interval. The original endpoint is $\lambda^\star = \mu + q\delta$, so
\begin{equation}
    \alpha_\infty = |f(\lambda^\star)| + (|f(\lambda_\eta^\star)| - |f(\lambda^\star)|),
\end{equation}
where $|f(\lambda^\star)| = \|f(\mathcal H)\|$ is the optimal subnormalization and $(|f(\lambda_\eta^\star)| - |f(\lambda^\star)|)$ is the excess. The excess is controlled by the distance between the original and enlarged endpoints, $|\lambda_\eta^\star-\lambda^\star| = \left(\frac{\eta}{2}-1\right)\delta$.  Applying the mean-value theorem gives
\begin{equation}
     \big||f(\lambda_\eta^\star)| - |f(\lambda^\star)|\big| \leq \left(\frac{\eta}{2}-1\right)\delta  \sup_{x\in \Omega_\eta^{\rm asin}}\left|f'(x)\right|,
     \label{eq:mean_value_theorem}
\end{equation}
where $\Omega_\eta^{\rm asin}$ is given in Eq.~\eqref{eq:omega_eta_asin}. Equation~\eqref{eq:mean_value_theorem} applies to any differentiable $f$ on the enlarged interval. Under the convex endpoint-reflection assumptions of Lemma~\ref{lem:convex_endpoint_reflection_saturation}, $|f'|$ is maximized at $\lambda_\eta^\star$, and hence
\begin{equation}
    \alpha_\infty = \|f(\mathcal H)\| + O([\eta-2]\delta |f'(\lambda_\eta^\star)|).
    \label{eq:arcsine_alpha_excess_big_o}
\end{equation}

\subsection{Resources for Sobolev regularization}
\label{sec:sobolev_resources}

\begin{proof}[Proof of Theorem~\ref{thm:sobolev_complexity}]
    Let $\Omega$ be the fitted spectral set, $C_\Omega$ be its Sobolev embedding constant, and $h$ be the periodic extension supplied by Theorem~\ref{thm:near_exponential_convergence}. Set $\epsilon_{\rm t}=\epsilon$, set 
    \begin{equation*}
        \sqrt{\gamma}=\frac{\epsilon}{2C_\Omega\|h\|_{H_r^w(\mathbb T)}},
    \end{equation*}
    and set $m$ sufficiently large such that
    \begin{equation*}
        C_2\exp[-C_1m^{1/\sigma}]\leq\frac{\epsilon}{2C_\Omega}.
    \end{equation*}
    Theorem~\ref{thm:near_exponential_convergence} then gives $\|g-\tilde{g}_m\|_{L^\infty(\Omega)}\leq\epsilon$. Moreover,
    \begin{equation*}
        \frac{C_2\exp[-C_1m^{1/\sigma}]}{\sqrt{\gamma}}\leq\|h\|_{H_r^w(\mathbb T)},
    \end{equation*}
    so Eq.~\eqref{eq:sobolev_coefficient_bound} gives
    \begin{equation}
        \alpha_m^{\rm reg}=\|\mathbf c\|_1\leq2C_wr^{-1/2}\|h\|_{H_r^w(\mathbb T)}\leq C_f\|f(\mathcal H)\|.
        \label{eq:regularized_alpha}
    \end{equation}
    The condition on $m$ requires
    \begin{equation}
        \begin{split}
            m = O\!\left(C_1^{-\sigma}\log^\sigma\!\left[\frac{2C_\Omega C_2}{\epsilon}\right]\right)= O\!\left(\log^\sigma[\epsilon^{-1}]\right),
        \end{split}
        \label{eq:regularized_modes_general}
    \end{equation}
    since $C_1$, $C_2$, and $C_\Omega$ are independent of $\epsilon$. Substituting Eqs.~\eqref{eq:regularized_alpha} and \eqref{eq:regularized_modes_general} into Eqs.~\eqref{eq:N_HS} and \eqref{eq:C_HS} gives
    \begin{equation*}
        N_{\rm HS}^{\rm reg}=O\!\left(\frac{C_f\|f(\mathcal H)\|}{\|f_m(\mathcal H)\ket{\psi}\|_2}\log\!\log[\epsilon^{-1}]\right),
    \end{equation*}
    which is Eq.~\eqref{eq:regularized_hs_calls}, and
    \begin{equation*}
        C_{\rm HS}^{\rm reg}=O\!\left(\frac{C_f\|f(\mathcal H)\|}{\|f_m(\mathcal H)\ket{\psi}\|_2}\frac{C_1^{-\sigma}}{\eta}s\log N\log^\sigma[\epsilon^{-1}]\right),
    \end{equation*}
    which is Eq.~\eqref{eq:regularized_complexity} for fixed $\eta>1$.
\end{proof}

For inversion, Section~\ref{sec:matrix_inversion} shows $C_f^{\rm inv}=O(1)$. It is therefore sufficient to take $m=O(\kappa\log^\sigma[\kappa\epsilon^{-1}])$. Together with Eq.~\eqref{eq:regularized_inverse_subnormalization}, Eqs.~\eqref{eq:N_HS} and \eqref{eq:C_HS} give
\begin{equation}
    N_{\rm HS}^{\rm reg,inv}=O\!\left(\kappa[\log\kappa+\log\!\log(\kappa\epsilon^{-1})]\right)
    \label{eq:regularized_inverse_hs_calls}
\end{equation}
calls to the Hamiltonian simulation algorithm, and
\begin{equation}
    C_{\rm HS}^{\rm reg,inv}=O\!\left(s\kappa^2\log^\sigma[\kappa\epsilon^{-1}]\log N\right)
    \label{eq:regularized_inverse_gate_complexity}
\end{equation}
two-qubit gates. The hidden constants may depend on the fixed $\sigma$, but are independent of $\kappa$, $m$, and $\epsilon$.

\section{Conclusions}
\label{sec:discussion}

We have introduced a Fourier-extension framework for constructing LCU block encodings of matrix functions. The central observation is that a Fourier-series approximation on a fitted spectral interval immediately induces an LCU decomposition of $g(x)$ through the functional calculus,
\begin{equation*}
    g_m(x)=\sum_{k=-m}^m c_k e^{ikx} \quad\longrightarrow\quad
    g_m(\mathcal G)=\sum_{k=-m}^m c_k e^{ik\mathcal G},
\end{equation*}
whenever $\mathcal G = \mathcal G^\dagger$. Fourier extensions approximate $g(x)$ on an extended domain, which allows exponential convergence even for non-periodic $g$, while redundancy provides an opportunity to minimize the coefficient $\ell^1$-norm. Since this $\ell^1$-norm is precisely the LCU subnormalization, Fourier extensions provide a direct mechanism for controlling both approximation error and postselection or amplitude amplification cost. Defining $\mathcal G$ and $g$ as in Eqs.~\eqref{eq:G_tau_definition} and \eqref{eq:g_definition} gives LCU decompositions of $f(\mathcal H)$, since $f(\mathcal H) = g(\mathcal G)$.

This gives a complementary route to matrix-function block encodings to the well-known QSVT framework~\cite{Gilyen2019}. Rather than assuming an input block encoding of $\mathcal H$ as in QSVT, we assume access to Hamiltonian simulation unitaries $e^{ik\tau\mathcal{H}}$. The same framework therefore applies naturally to eigenvalue transforms of Hermitian matrices and, after Hermitian dilation, to odd singular-value transforms of general matrices discussed surrounding Eq.~\eqref{eq:hermitian_dilation}.

The three coefficient-selection strategies developed in this work represent different points in the trade-off between convergence rate, subnormalization, and versatility:
\begin{enumerate}
    \item Reflected extensions: By reflecting the function at the endpoint of the approximation interval, we obtain Fourier extensions whose coefficients saturate the block-encoding norm lower bound (for convex $f$). This gives block encodings with optimal subnormalization, although with just first-order algebraic convergence,
    \begin{equation*}
        \alpha_m = \|f(\mathcal H)\|-\epsilon_m,\qquad \epsilon_m = O(m^{-1}).
    \end{equation*}
    
    \item Arcsine-Taylor extensions: By truncating the Taylor series of $g(\arcsin z)$ then using a change of variables $z=\sin x$, we obtain an explicit Fourier representation with bounded subnormalization and exponential convergence on $[-\pi/\eta,\pi/\eta]$. For the identity function this gives
    \begin{equation*}
        \epsilon_m = O\!\left(m^{-3/2}\sin^m\!\left[\frac{\pi}{\eta}\right]\right).
    \end{equation*}
     This construction is especially useful for entire functions where $\eta$ is unrestricted, and provides a closed-form alternative to numerically computed coefficients. For non-entire functions, such as square roots or inverses, $\eta$ must be chosen so that the enlarged analytic region avoids branch points or poles.
     
    \item Sobolev-regularized extensions: By treating the Fourier-extension coefficients as the solution of a regularized approximation problem, one can exploit the redundancy of the restricted Fourier frame. Using the $H^1$-norm for fitting and the $H^1_r$-norm for regularization controls both the approximation in $L^\infty$ and the coefficient $\ell^1$-norm. For analytic functions, the regularized construction achieves near-exponential convergence to a prescribed target tolerance for every fixed $\sigma>1$,
    \begin{equation*}
        \epsilon_m=O\!\left(\exp[-C_1m^{1/\sigma}]+\epsilon_{\rm t}\right),
    \end{equation*}
    while maintaining a subnormalization bound that is independent of the target tolerance. Section~\ref{sec:application_to_various_functions} shows the subnormalization scales optimally for the identity, exponential, and inverse functions, yet has an $O(\kappa^{1/4})$ overhead above the optimal value for the square-root function. The method can be applied to general fitted spectral sets, including disconnected intervals, which is essential for indefinite matrix inversion.

\end{enumerate}

Fourier extensions provide a system for producing matrix-function block encodings from Hamiltonian simulations with bounded subnormalization. Future work should address developing the framework into practical end-to-end quantum algorithms for specific applications. This requires efficient Hamiltonian simulation implementations for a given application, so that the application-specific cost of implementing the unitaries $e^{\pm ik\tau\mathcal{H}}$ can be incorporated into the resource analysis. There is also scope for further improvements to the coefficient-selection problem to prioritize other points on the error-subnormalization-versatility trilemma, and to specialize towards particular matrix functions and problem classes that exploit additional spectral or structural information about the problem. The most promising applications are likely to be those where the approximation problem and the Hamiltonian-simulation implementation can be designed together.

\begin{acknowledgments}
    PB is supported by The University of Manchester through the Dame Kathleen Ollerenshaw Fellowship. BA is supported by the Natural Sciences and Engineering Research Council of Canada (NSERC) through grant RGPIN/2026-04531.
\end{acknowledgments}

\bibliography{apssamp}

\end{document}